\documentclass[twocolumn,showpacs,preprintnumbers,amsmath,amssymb]{revtex4}

\usepackage{graphicx}
\usepackage{dcolumn}
\usepackage{bm}
\usepackage{epsfig}
\usepackage{rotating}
\usepackage{subfigure}

\begin{document}


\preprint{APS/000-CSP}

\title{Driven Diffusive Systems: How Steady States Depend on Dynamics}

\author{Wooseop Kwak}
\email{wkwak@hal.physast.uga.edu}
\affiliation{%
Center for Simulational Physics, The University of Georgia, Athens, GA, 30602-2451, USA\\
}%

\author{D. P. Landau}%
\email{dlandau@hal.physast.uga.edu}
\affiliation{%
Center for Simulational Physics, The University of Georgia, Athens, GA, 30602-2451, USA\\
}%

\author{B. Schmittmann}
\email{schmittm@vt.edu}
\affiliation{%
Center for Stochastic Processes in Science and Engineering,\\
Physics Department, Virginia Tech, Blacksburg, VA 24061-0435, USA\\
}%

\date{February 26, 2004}

\begin{abstract}
In contrast to equilibrium systems, non-equilibrium steady states 
depend explicitly on the underlying dynamics. Using Monte Carlo 
simulations with Metropolis, Glauber and heat bath rates, we 
illustrate this expectation for an Ising lattice gas, driven 
far from equilibrium by an `electric' field. 
While heat bath and Glauber rates generate essentially identical data
for structure factors and two-point correlations, Metropolis rates 
give noticeably weaker correlations, as if the `effective' temperature
were higher in the latter case. We also measure energy histograms
and define a simple ratio which is exactly known and closely related
to the Boltzmann factor for the equilibrium case. For the driven
system, the ratio probes a thermodynamic derivative which is found to
be dependent on dynamics. 
\end{abstract}

\pacs{05.20.-y, 05.70.Ln, 64.60.Cn, 66.30.Dn}

\maketitle

\section{\label{sec:level1}Introduction\newline
}

In statistical physics, Monte Carlo (MC) simulations play a major role for
the study of phase transitions and critical phenomena, as well as ordered
and disordered phases \cite{MC}. Leaving out many details of the art of
computing, the broad outline of the simulation process is easily summarized.
For systems both in and far from thermal equilibrium, dynamic rates 
$ W[\sigma \rightarrow \sigma ^{\prime }]$ are defined which specify how a
given configuration $\sigma $ evolves into a new one, $\sigma ^{\prime }$,
when an update is attempted.
The simulations then generate long sequences of such configurations. Once
initial transients have decayed, \emph{time-independent} (stationary)
observables -- which will be our focus in the following -- can be computed 
as configurational averages. For a system in thermal equilibrium, characterized
by a Hamiltonian $\mathcal{H}$, it is well known that \emph{any}
choice of $W$'s, as long as they satisfy detailed balance with respect to $H$,
will generate configurations distributed according to the \emph{same}
Boltzmann factor, $\exp (-\beta H)$. In other words, time-independent
observables, including both universal and non-universal properties, are
independent of the choice of rates, provided detailed balance holds.
The resulting freedom can be exploited to design particularly efficient
codes, such as cluster algorithms \cite{SW}. In stark contrast, no such
`decoupling' of dynamic and stationary characteristics occurs for systems
driven out of equilibrium: even though non-equilibrium steady states (NESS)
display time-independent observables, these are sensitive to
modifications of the dynamic rates. This behavior can be traced back to the
violation of detailed balance which is an inherent feature of
non-equilibrium systems \cite{SZ,revs}.

While the sensitivity of NESS\ to the choice of the dynamics has been noted
before \cite{KLS,Krug}, no systematic computational study has yet been
undertaken. In this paper, we consider two models: the standard Ising
lattice gas and its non-equilibrium cousin, the driven Ising lattice gas (or
KLS model, after the initials of its inventors \cite{KLS}). Both involve
particles diffusing on a lattice, subject to an excluded volume constraint
and an (attractive) nearest-neighbor interaction. The total number of
particles remains conserved. For the Ising lattice gas, the rates for
particle hops to unoccupied nearest-neighbor sites are chosen to satisfy
detailed balance, with respect to the Ising Hamiltonian. In contrast, the
driven version involves an additional external force which acts on the
particles much like an electric field on (positive) charges: aligned with a
lattice axis (e.g., $y$), it favors particle hops along its direction. In
conjunction with periodic boundary conditions, this bias breaks detailed
balance and establishes a \emph{non-equilibrium} steady state. This NESS
differs drastically from its equilibrium counterpart, exhibiting generic
long-range correlations, a novel universality class, and highly anisotropic
ordered phases \cite{SZ}.

To illustrate the importance of the transition rates for the NESS, we
measure structure factors and two-point spatial correlations in the driven
case, using Metropolis \cite{Met}, Glauber \cite{glauber}, and heat bath 
\cite{heatbath} rates. To date, simulations of the driven model have focused
almost exclusively on\ Metropolis rates \cite{VM}; other rates have only
been invoked in some analytic studies \cite{Krug,ana}. While the first two
are easily implemented for conserved particle number, an appropriate
generalization of heat bath rates is designed here for the first time. To
avoid complications due to inhomogeneous ordered phases, we choose
temperatures above or at criticality. For comparison, we also show the same
quantities for the (equilibrium) Ising model: as expected, they are found to
be identical up to statistical fluctuations, and almost perfectly isotropic.
In stark contrast, when the drive is turned on, we find strongly anisotropic
behavior and markedly different values for the three choices of rates. While
all driven cases exhibit the signatures of long-ranged decay in real space,
the correlations are much weaker for Metropolis rates than for either
Glauber or heat bath ones.

As an additional probe into the differences between the standard Ising
lattice gas and its driven cousin, we construct energy histograms (with
respect to the Ising Hamiltonian) for both. For the equilibrium case, these
are of course intimately related to the Boltzmann distribution and contain a
wealth of thermodynamic information. For the driven system, they are easily
measured in a simulation, but their physical interpretation has not been
established yet. Here, we consider a simple histogram ratio whose
equilibrium limit is easily derived, and compute its non-equilibrium
counterpart. 

This paper is organized as follows. We first introduce our models and the
three types of dynamics, followed by a brief discussion of our key
observables. We then present our data for two-point correlations, structure
factors and energy histograms. We conclude with a summary, offering a
conjecture for the origin of the observed differences between the three rate
functions.

\section{\label{sec:level2}Background\newline
}

\subsection{The models}

In this section, we introduce our two prototype models, namely, the Ising
lattice gas and its driven version. Both are defined on an $M\times L$
square lattice in two dimensions, with fully periodic boundary conditions.
Each site $\mathbf{i}$ is either occupied by a particle or empty, which we
denote by a spin variable $\mathbf{\sigma _{i}}$ taking two values: $+1$
(occupied) or $-1$ (empty). For the equilibrium Ising model, we can specify a
(global) Hamiltonian:

\begin{equation}
\mathcal{H}[\sigma ]=-J\sum_{<\mathbf{i},\mathbf{j}>}\mathbf{\sigma
_{i}\,\sigma _{j}},  \label{Ising}
\end{equation}%
where the sum runs over nearest-neighbor pairs of sites, and $J>0$ denotes
the binding energy. In order to access the Ising critical point, we consider
only half-filled systems: $(LM)^{-1}\sum_{\mathbf{i}}\sigma \mathbf{_{i}}=0$%
. When coupled to a heat bath at temperature $T$, the probability, $%
P_{0}(\sigma )$, to find the system in configuration $\sigma $ is
controlled by the well-known Boltzmann factor: $P_{0}(\sigma )\propto \exp
(-\beta H)$ with $\beta =1/k_{B}T$. Here and in the following, the 
subscript $0$ will always denote equilibrium quantities.

The usual technique for simulating such a distribution is to introduce a
dynamics in configuration space. We choose a suitable set of transition
rates, $W[\sigma \rightarrow \sigma ^{\prime }]$, which specify how a
configuration $\sigma $ evolves into a new one, $\sigma ^{\prime }$, in unit
time. For simplicity, we only consider transitions in which $\sigma $ and $%
\sigma ^{\prime }$ differ by a \emph{single} nearest-neighbor particle-hole
exchange. Now, the probability distribution, $P(\sigma ,t)$, becomes
time-dependent and satisfies a master equation (written, for simplicity, in
continuous time): 
\begin{equation}
\partial _{t}P(\sigma ,t)=\sum_{\sigma ^{\prime }}\left\{ W[\sigma ^{\prime
}\rightarrow \sigma ]P(\sigma ^{\prime },t)-W[\sigma \rightarrow \sigma
^{\prime }]P(\sigma ,t)\right\}  \label{master}
\end{equation}%
Its stationary solution, $P(\sigma )\equiv \lim_{t\rightarrow \infty
}P(\sigma ,t)$, controls all time-\emph{independent} properties. It is
unique, under fairly generic conditions on the rates. To ensure that the
desired equilibrium distribution, $P_{0}(\sigma )$, is reproduced, one
chooses $W$'s which satisfy the detailed balance condition: 
\begin{equation}
\frac{W[\sigma \rightarrow \sigma ^{\prime }]}{W[\sigma ^{\prime
}\rightarrow \sigma ]}=\frac{P_{0}(\sigma ^{\prime })}{P_{0}(\sigma
)}=e^{-\beta \left\{ \mathcal{H}[\sigma ^{\prime }]\mathcal{-H}[\sigma
]\right\} }  \label{det_bal_1}
\end{equation}%
Of course, this just ensures that every $\left\{ ...\right\} $ bracket in
Eq. (\ref{master}) vanishes. An important quantity which enters here is the
energy difference of two configurations, $\sigma ^{\prime }$ and $\sigma $: 
\begin{equation}
\Delta _{0}\equiv \mathcal{H}[\sigma ^{\prime }]\mathcal{-H}[\sigma ]
\label{Del_H}
\end{equation}%
A simple way of satisfying detailed balance is to impose rates which are
functions of this difference alone: $W[\sigma \rightarrow \sigma ^{\prime
}]\equiv w(\beta \Delta _{0})$ where the function $w$ must satisfy  
\begin{equation}
w(x)=w(-x)e^{-x}\quad ,  \label{det_bal_2}
\end{equation}%
by virtue of Eq.~(\ref{det_bal_1}) but is otherwise arbitrary. All three
rate functions to be considered below -- Metropolis, Glauber, and heat bath
-- are constructed in this way, but differ in some important details.

An obvious way of driving a system into a \emph{non-equilibrium} steady
state is to impose rates that \emph{violate} detailed balance. A prototype
model that has attracted much interest due to its remarkable properties is
the driven Ising lattice gas (or KLS model) \cite{KLS,SZ}. It differs from
the standard Ising model through the presence of an external force $\mathcal{%
E}$, aligned with a particular lattice axis (the $y$-direction). When
a particle attempts to jump to an empty nearest-neighbor site, it is not
only affected by the local energetics, incorporated in 
Eq.~(\ref{Ising}), but also by the drive: similar to an
electric field, $\mathcal{E}$ favors (suppresses) particle hops along
(against) the selected direction, leaving transverse exchanges unaffected. A
straightforward extension of Eq.~(\ref{Del_H}) is to include the work done
by the field, i.e., to define a \emph{local} `energy' difference of the form 
\begin{equation}
\Delta \equiv \mathcal{H}[\sigma ^{\prime }]\mathcal{-H}%
[\sigma ]-\epsilon \mathcal{E}  \label{Del_HE}
\end{equation}%
Here, $\epsilon =0$ for two configurations differing only by a transverse
jump, and $\epsilon =+1$ $(-1)$ if the particle hops along (against) the
field in the move. We can now choose rates of the form (\ref{det_bal_2})
with $x=\beta \Delta $. However, it is essential to note that
the \emph{combination} of uniform drive and periodic boundary conditions 
\emph{precludes} the existence of a \emph{global} Hamiltonian for the driven
system. A unique steady state, $P(\sigma )$, establishes itself but
cannot be expressed in terms of a Boltzmann factor. To maximize the
non-equilibrium effects, we choose infinite $\mathcal{E}$ for our
simulations, i.e., a particle will never jump against the field.

\subsection{Three different rate functions.}

In this subsection, we introduce the three choices of transition rates --
Metropolis, Glauber, and heat bath -- which will be compared in the
following. For the first two choices, the relevant quantity is the local
`energy' difference between the \emph{final }($\sigma ^{\prime }$) and the 
\emph{initial} ($\sigma $) configuration. For the third choice, the rate is
independent of the initial configuration; instead, the selection criterion
involves the local energy difference of the two possible \emph{final}
configurations, $\sigma $ and $\sigma ^{\prime }$. For the equilibrium Ising
model, energy differences are easily computed from Eq.~(\ref{Del_H}), and
each rate satisfies the detailed balance condition; for the driven model, we
invoke Eq.~(\ref{Del_HE}), and detailed balance is violated. Random numbers are
selected uniformly from the interval $(0,1)$.

\emph{Metropolis dynamics.} For this choice of rates, we randomly select a
nearest-neighbor pair $\mathbf{i}$, $\mathbf{j}$ of sites with different
occupanices, $\sigma _{\mathbf{i}}\neq \sigma _{\mathbf{j}}$. We denote the
original configuration as $\sigma $ and let $\sigma ^{\prime }$ be the
configuration with $\sigma _{\mathbf{i}}$, $\sigma _{\mathbf{j}}$ switched
(i.e., $\sigma _{\mathbf{i}}^{\prime }=\sigma _{\mathbf{j}}$; $\sigma _{%
\mathbf{j}}^{\prime }=\sigma _{\mathbf{i}}$). The transition from $\sigma $
to $\sigma ^{\prime }$ is controlled by the Metropolis rate function, $%
w_{Met}(x)=\min (1,e^{-x})$. To be specific, we first compute $x=\beta
\Delta$. If $x\leq 0$, the attempt (exchange) is accepted;
if, however, $x>0$, we draw a random number $z$ and perform the exchange
only if $z\leq e^{-x}$. Clearly, energetically favorable moves are always
performed while only a fraction of costly ones is accepted. As temperature
increases, this fraction approaches $1$ in a monotonic fashion.

\emph{Glauber dynamics.} Similar to Metropolis dynamics, the implementation
of this algorithm involves, first, selecting two nearest-neighbor sites with
different occupancies. Again, $\sigma ^{\prime }$ refers to the
configuration with switched occupancies. Exchanges are then controlled by
the Glauber rate function, $w_{Gl}(x)=1/(1+e^{x})$. Again, we compute $%
x=\beta \Delta$ and draw a random number, $z$. If $z\leq
1/(1+e^{x})$, we accept the exchange; otherwise, it is rejected. While
energetically favorable moves are not necessarily accepted, they are always
more probable than unfavorable ones.

\emph{Heat bath dynamics.} As pointed out above, the interpretation of $%
\sigma $ and $\sigma ^{\prime }$ is different here: These refer to the two
possible final configurations of the central particle-hole pair. Showing
only its local neighborhood in the lattice, we define 

\begin{equation}
\sigma \equiv 
\begin{tabular}{|l|l|l|l|}
\hline
& $\sigma _{3}$ & $\sigma _{4}$ &  \\ \hline
$\sigma _{2}$ & $+1$ & $-1$ & $\sigma _{5}$ \\ \hline
& $\sigma _{1}$ & $\sigma _{6}$ &  \\ \hline
\end{tabular}%
\text{ and }\sigma ^{\prime }\equiv 
\begin{tabular}{|l|l|l|l|}
\hline
& $\sigma _{3}$ & $\sigma _{4}$ &  \\ \hline
$\sigma _{2}$ & $-1$ & $+1$ & $\sigma _{5}$ \\ \hline
& $\sigma _{1}$ & $\sigma _{6}$ &  \\ \hline
\end{tabular}
\label{sigma-x}
\end{equation}%
for bonds along the $x$-axis, and 
\begin{equation}
\sigma \equiv 
\begin{tabular}{|l|l|l|}
\hline
& $\sigma _{2}$ &  \\ \hline
$\sigma _{1}$ & $+1$ & $\sigma _{3}$ \\ \hline
$\sigma _{6}$ & $-1$ & $\sigma _{4}$ \\ \hline
& $\sigma _{5}$ &  \\ \hline
\end{tabular}%
\text{ and }\sigma ^{\prime }\equiv 
\begin{tabular}{|l|l|l|}
\hline
& $\sigma _{2}$ &  \\ \hline
$\sigma _{1}$ & $-1$ & $\sigma _{3}$ \\ \hline
$\sigma _{6}$ & $+1$ & $\sigma _{4}$ \\ \hline
& $\sigma _{5}$ &  \\ \hline
\end{tabular}
\label{sigma-y}
\end{equation}%
for bonds along $y$; i.e., parallel to the drive. We also define 
\begin{equation}
h\equiv \sum_{\mathbf{i}=1}^{3}\sigma _{\mathbf{i}}-\sum_{\mathbf{i}%
=4}^{6}\sigma _{\mathbf{i}}  \label{field}
\end{equation}%
In equilibrium, the heat bath algorithm is of course isotropic: For both
types of bonds, we select configuration $\sigma $ if a random number $z$ 
satisfies $z\leq 1/(1+e^{-2\beta h})$;
otherwise, we choose $\sigma ^{\prime }$. For the driven case, this rule is
only applied to bonds transverse to the drive; for parallel bonds, at
infinite $\mathcal{E}$, we choose configuration $\sigma $ with probability $%
1 $.

\begin{figure}[tbph]
\begin{center}
\mbox{ 
  {\scalebox{0.45}
	  {\epsfig{file=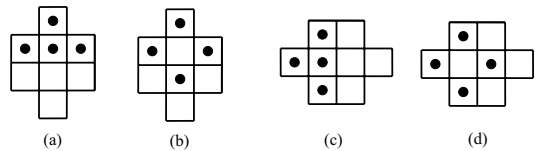, bbllx=60pt,bburx=680pt, bblly=730pt, bbury=550pt, angle=-0 }}
  }
} 
\end{center}
\par
\vspace{18mm}
\caption{A central pair and a particular configuration of its six nearest neighbors. 
Occupied sites are indicated by solid circles.} 
\end{figure}

To appreciate the commonalities and differences of the three algorithms, it
is useful to consider a simple example with infinite drive. Fig.~1 shows a
central pair and a particular configuration of its six nearest neighbors. If the pair 
is aligned with the field direction (Figs.~1a, b), all three dynamics 
generate the same outcome: each will select Fig.~1a as the final configuration 
with probability $1$ for any value of $\beta J$.

This is not the case for bonds \emph{transverse} to the field (Figs.~1c, d).
For the purposes of this argument, we choose $\beta J=0.1$. We denote the
configuration shown in Fig.~1c (d) by $\sigma $ ($\sigma ^{\prime }$). The
`energy' difference $\Delta =12J$ is easily computed 
from Eq.~(\ref{Del_H}) or (\ref{Del_HE}). Given a random number $z$,
the Metropolis algorithm will accept a transition from $\sigma $
to $\sigma ^{\prime }$ only if $z\leq $ $e^{-12\beta J}\simeq 0.30$, 
while the reverse transition ($\sigma ^{\prime }$
to $\sigma $) is always accepted. For Glauber dynamics, the transition from $%
\sigma $ to $\sigma ^{\prime }$ is accepted if $z$ $\leq 1/(1+e^{12\beta J})$
$\simeq $ $0.23$, while the reverse transition is accepted if $z$ $\leq
1/(1+e^{-12\beta J})\simeq 0.77$. Finally, the heat bath algorithm will
choose $\sigma $ as the final state if $z$ $\leq 1/(1+e^{-12\beta J})\simeq $
$0.77$, and $\sigma ^{\prime }$ otherwise.

The notable differences are these: First, the Metropolis algorithm accepts
unfavorable moves with higher probability than either heat bath or Glauber: 
$e^{-x}\geq $ $1/(1+e^{x})$. As a result, it is more likely to explore
unphysical domains of configuration space. Yet, it also accepts favorable
moves with higher probability, and thus leads to a more active dynamics. 
Comparing heat bath and Glauber rates, we note that both
subdivide the unit interval into the same subsections ($0.23$ vs $0.77$).
Hence, they generate \emph{statistically} very similar trajectories in
configuration space. Update by update, however, the trajectories can differ:
if, e.g., the initial configuration is $\sigma $ and the random number
turns out to be $0.1$, the heat bath algorithm will choose $\sigma ^{\prime
} $ as the final configuration, while the Glauber rule leads to an
exchange since $z<0.23$. Yet, we will see below that this subtle 
difference does not affect the data.  

\subsection{Structure factors and two-point correlations.}

Below their critical temperatures, both the Ising lattice gas and the KLS
model phase-segregate into regions of high and low density, by virtue of the
conservation law on the number of particles. Typical low-temperature
configurations, for both models, show a single strip of high-density phase
and its low-density mirror image. For the Ising model, the strip
orients itself such as to minimize the energetic cost of interfacial length.
In contrast, the low-temperature
strip of the KLS model is \emph{always} aligned with the direction of the
drive, and the minimization of interfacial length does not play a dominant
role (cf.~Fig.~2). A quantity which easily distinguishes disordered
configurations from such inhomogeneous ones is the (equal-time) structure
factor. Written in spin language, it is defined as 
\begin{equation}
S(\mathbf{k})=\frac{1}{ML}\left\langle \left| \sum_{\mathbf{j}}e^{i\mathbf{k}%
\cdot \mathbf{j}}\sigma _{\mathbf{j}}\right| ^{2}\right\rangle .
\label{eq:sfactor}
\end{equation}%
Here, $\mathbf{k}$ is a wave vector, taking discrete values $\mathbf{k}%
=(2\pi n_{x}/M,2\pi n_{y}/L)$ with $n_{x}=0,1,...,M-1$ and $%
n_{y}=0,1,...,L-1 $. For simplicity, we write $S(n_{x},n_{y})$ in the
following, and use $S(1,0)$ and $S(0,1)$ to detect strips aligned with the $y$-
or $x$-axis, respectively. 
For a perfectly ordered strip aligned with $y$, $%
S(1,0)\simeq 0.41ML$ is maximized; in contrast, a disordered configuration
results in $S(1,0)=O(1)$. 
Further, the structure factor is the Fourier transform
of the two-point correlation function, $G(\mathbf{r})$, defined via 
\begin{equation}
G\left( \mathbf{r}\right) \equiv \left\langle \sigma _{\mathbf{0}}\sigma _{%
\mathbf{r}}\right\rangle -\left\langle \sigma _{\mathbf{0}}\right\rangle
\left\langle \sigma _{\mathbf{r}}\right\rangle  \label{eq:paircorrelationfn}
\end{equation}%
We assume translational invariance (modulo the lattice size) and invoke the
half-filling constraint, whence $\left\langle \sigma _{\mathbf{r}%
}\right\rangle =\left\langle \sigma _{\mathbf{0}}\right\rangle =0$. The same 
constraint imposes the sum
rule $\sum_{\mathbf{r}}G\left( \mathbf{r}\right)=S(\mathbf{0})= 0$. 
Hence, negative values
of $G\left( \mathbf{r}\right) $ for certain values of the argument should
not come as a surprise.

\subsection{Energy histograms.}

For both the equilibrium Ising model and its driven counterpart, it is
straightforward to accumulate a (normalized) energy histogram $H(E,\beta )$, with
respect to the energy function defined in Eq.~(\ref{Ising}). 
For the equilibrium Ising model, 
$H_{0}(E,\beta )$ is intimately related to the Boltzmann distribution: 
if $W\left( E\right) $ denotes the density of states and $Z(\beta )$ 
the canonical
partition function, we have

\begin{equation}
H_{0}\left( E,\beta \right) = Z^{-1}(\beta )W\left( E\right) e^{-\beta E}  
\label{Eq-hist}
\end{equation}
Clearly, the right hand side is the probability $P_{0}(E,\beta )$ to find
the system with energy $E$. The power of the histogram method \cite{hist}
resides in the observation that, up to statistical errors, a single
histogram measured at temperature $1/\beta $ is sufficient to construct $%
P_{0}(E,\beta ^{\prime })$ at all other temperatures $1/\beta ^{\prime }$: 
\begin{equation}
P_{0}(E,\beta ^{\prime })=\frac{H_{0}\left( E,\beta \right) e^{-(\beta
^{\prime }-\beta )E}}{\sum_{E^{\prime }}H_{0}\left( E^{\prime },\beta
\right) e^{-(\beta ^{\prime }-\beta )E^{\prime }}}  \label{Eq-P(E,T)}
\end{equation}
This allows us to compute the moments of the energy distribution as
functions of temperature and extract a wealth of thermodynamic information.

For the driven lattice gas, $H(E,\beta )$ is easily compiled in a
simulation. However, Eq.~(\ref{Eq-hist}) certainly does not hold for the
non-equilibrium steady state. In particular, exact solutions of small
systems \cite{small} demonstrate unambiguously that, at a given temperature,
configurations with the same energy need not have the same probability. At
best, we can write, using the Kronecker symbol, 
\begin{equation}
H\left( E,\beta \right) = \sum_{\sigma
}\delta _{E,\mathcal{H}[\sigma ]}P(\sigma )\equiv \exp [-F(E,\beta
)]  \label{Eq-KLS-H}
\end{equation}
where $F(E,\beta )$ is an as yet unknown function of its variables which
will certainly depend on the chosen dynamics. 

In the following, we probe $F(E,\beta )$ by considering a very simple
ratio: We measure two histograms at different inverse temperatures, $\beta
_{1}$ and $\beta _{2}$ and construct

\begin{equation}
R(E,E^{\prime })\equiv \frac{H(E,\beta _{1})}{H(E^{\prime },\beta _{1})}%
\times \frac{H(E^{\prime },\beta _{2})}{H(E,\beta _{2})}
\label{eq:historatio}
\end{equation}
for a range of $E,E^{\prime }$. In equilibrium,
this ratio is just a simple exponential: $R_{0}(E,E^{\prime })=\exp [-(\beta
_{1}-\beta _{2})(E-E^{\prime })]$, since all normalization factors
cancel. For the driven system, little is known except 
\begin{equation}
R(E,E^{\prime })\equiv \exp [F(E,\beta _{2})-F(E,\beta
_{1})-F(E^{\prime },\beta _{2})+F(E^{\prime },\beta _{1})].
\label{dds-historatio}
\end{equation}
This form will be analyzed further below.

To conclude this section, we establish a few conventions and summarize the
technical details of the simulations. All temperatures in the following are
quoted in units of $J/k_{B}$; an important reference point is the Onsager
temperature $T_{0}=-2/\ln (\sqrt{2}-1)\simeq 2.269$ \cite{Onsager} which
marks the critical point of the two-dimensional Ising model. The equilibrium
lattice gas and the driven system differ only in one parameter: $\mathcal{E}%
=0$ vs $\mathcal{E}=1000$, respectively. Such a large value for 
$\mathcal{E}$ suppresses (almost) all moves against the drive, and is therefore
effectively infinite. When a quantity, e.g., the critical temperature for
the driven system, has been measured in different dynamics, we will use
superscripts M (Metropolis), H (heat bath), and G (Glauber) to distinguish
them, as in $T_{c}^{M}$, $T_{c}^{H}$, and $T_{c}^{G}$. The data for
structure factors and two-point correlations were obtained on $100\times 100$
systems while the histogram simulations used a smaller system size, 
$40\times 40$. In each case, $1$ MCS corresponds to one update attempt per
site on average. For the larger system, each run lasted $2\times 10^{6}$ MCS.
The first $10^{6}$ MCS were discarded to ensure that the system had reached
steady state, and data were taken every $100$ MCS over the second half of
the run. For better statistics, $20$ independent runs were performed and
averaged. For the smaller size, $4\times 10^{6}$ MCS 
were discarded, followed by $12\times 10^{6}$ measurements. 

\begin{widetext}

\section{\label{sec:level3}Results\newline
}

\begin{figure}[tbph]
\begin{center}
\mbox{ 
	  {\scalebox{0.45}
	      {\epsfig{file=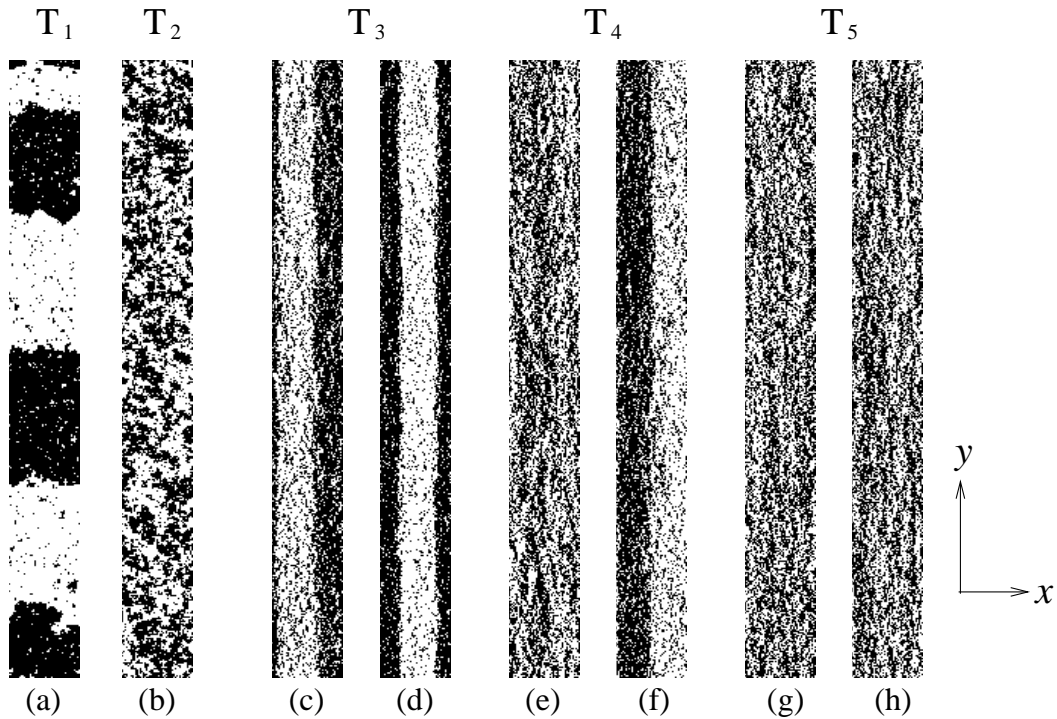, bbllx=80,bburx=680, bblly=0, bbury=550, angle=-0}}
	     	  }
    }
\end{center}
\par
\vspace{-7mm}
\caption{Typical configurations on a $48\times 432$ lattice for the equilibrium
Ising model using heat bath dynamics at $T_{1}=2.00$ (a) and $T_{2}=2.80$ (b),
and for its driven cousin at three temperatures: $T_{3}=2.90$ (c, d), 
$T_{4}=3.30$ (e, f) and $T_{5}=3.70$ (g, h). The first (second) configuration at  
each temperature was obtained using Metropolis (heat bath) dynamics. 
In each simulation, the data were collected after discarding 
$2\times 10^{7}$ MCS for the equilibrium system  
and $10^{7}$ MCS for its driven counterpart.}
\label{fig:strip}
\end{figure}

\begin{figure}[tbph]
\begin{center}
\mbox{ 
	  {\scalebox{0.19}
	     {\epsfig{file=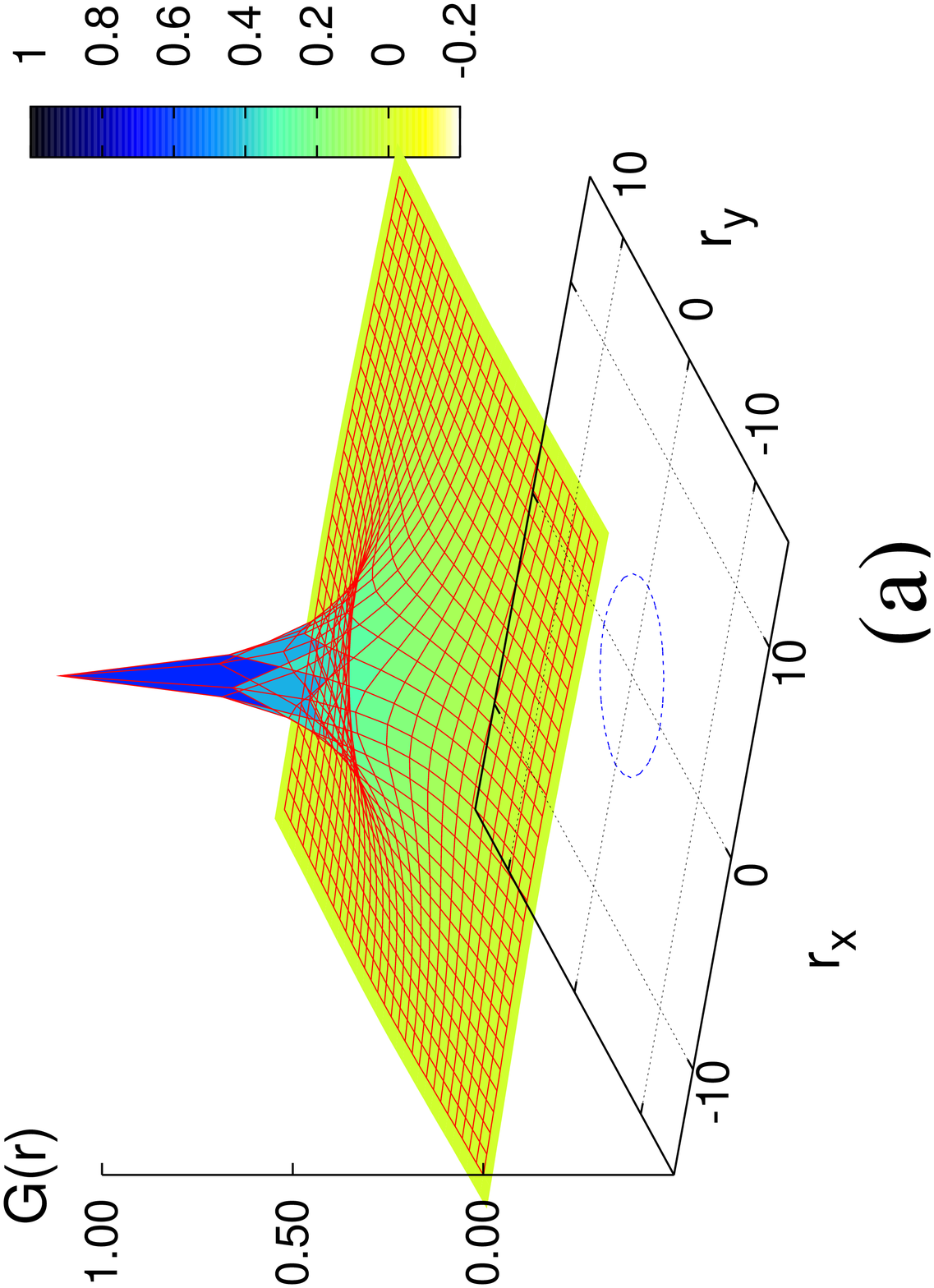, angle=-90}}
	  }
	  {\scalebox{0.19}
	     {\epsfig{file=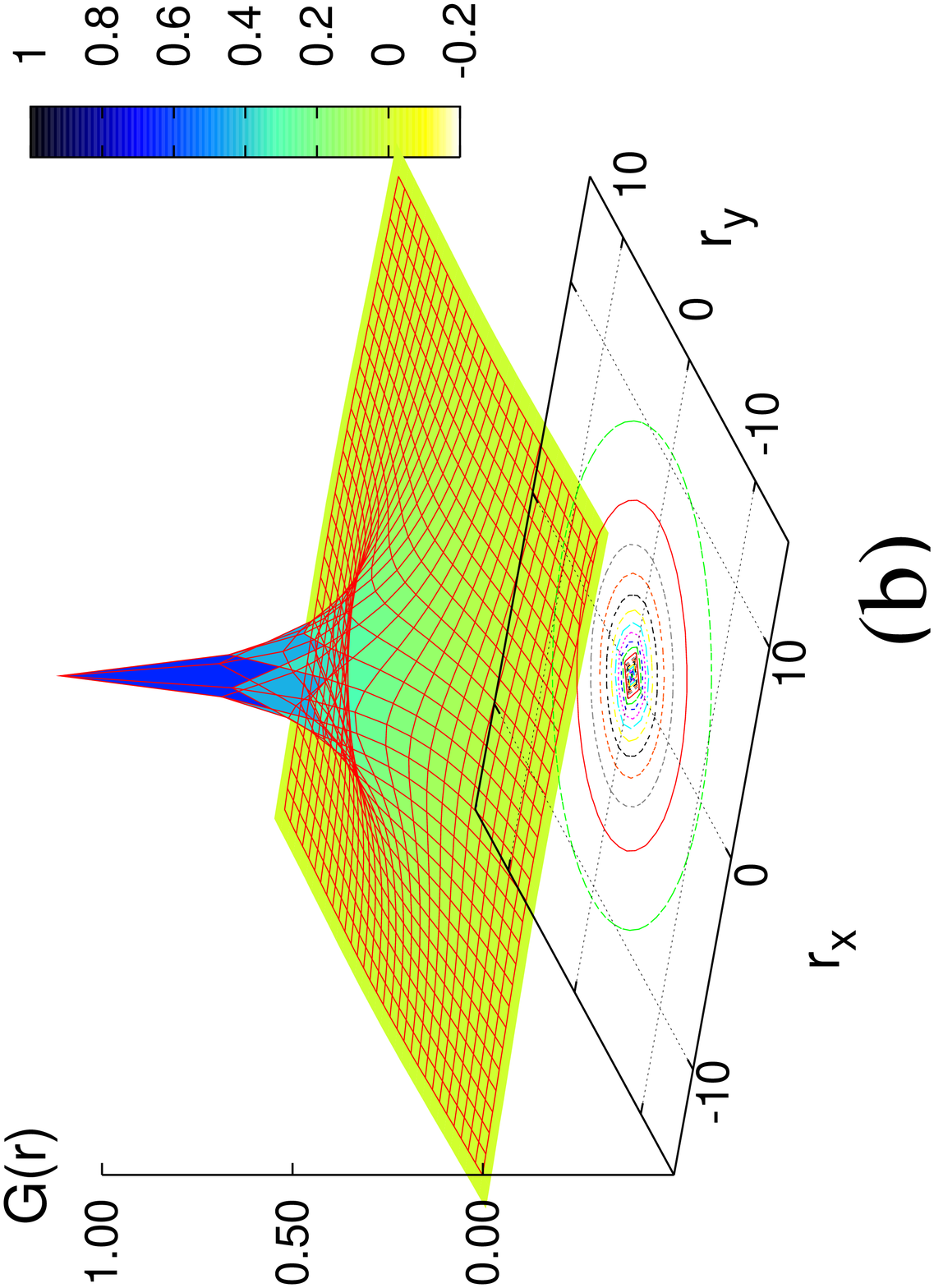, angle=-90}}
	  }
	  {\scalebox{0.19}
	     {\epsfig{file=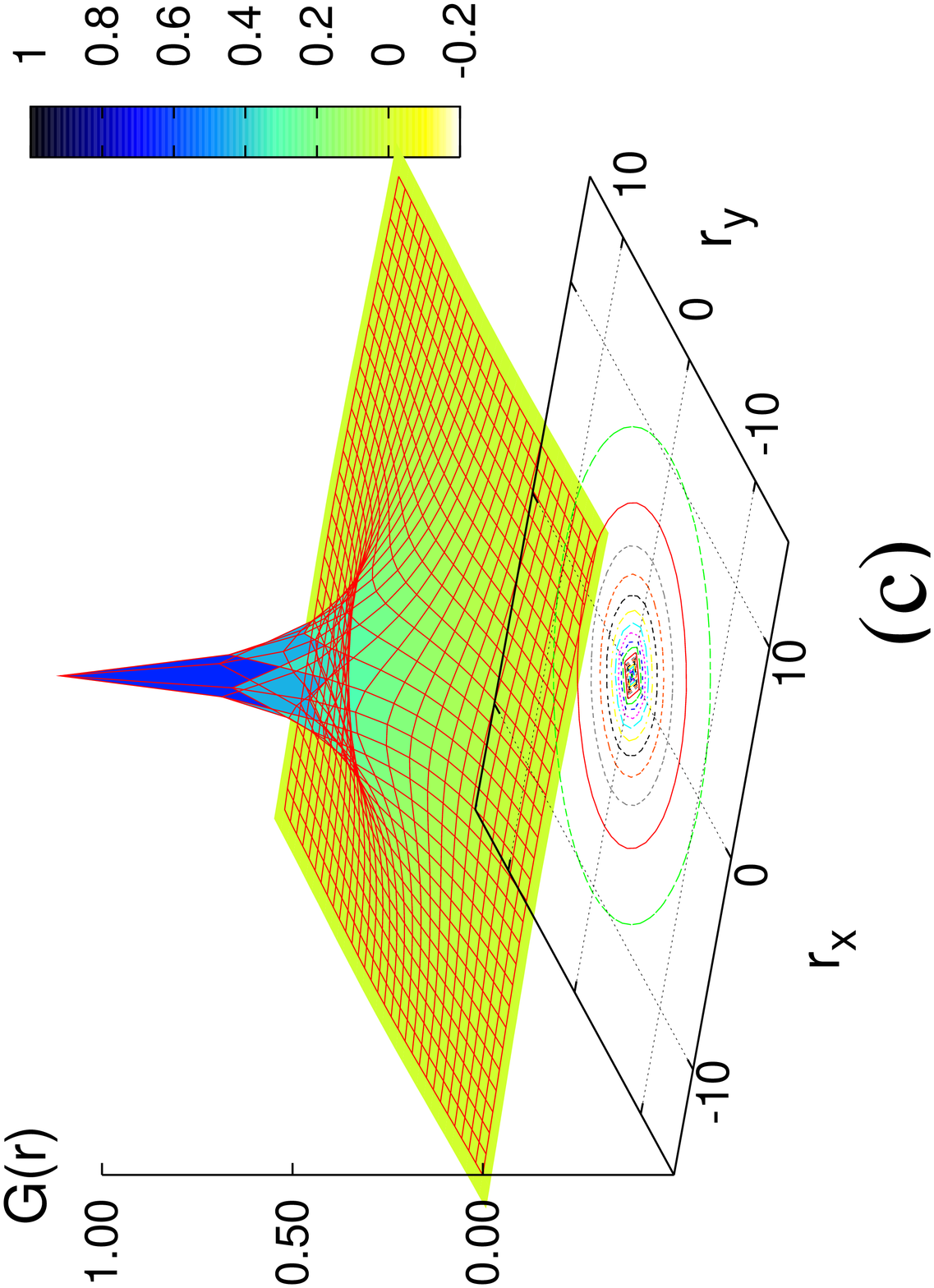, angle=-90}}
	  }
    } 
\mbox{ 
	  {\scalebox{0.19}
	     {\epsfig{file=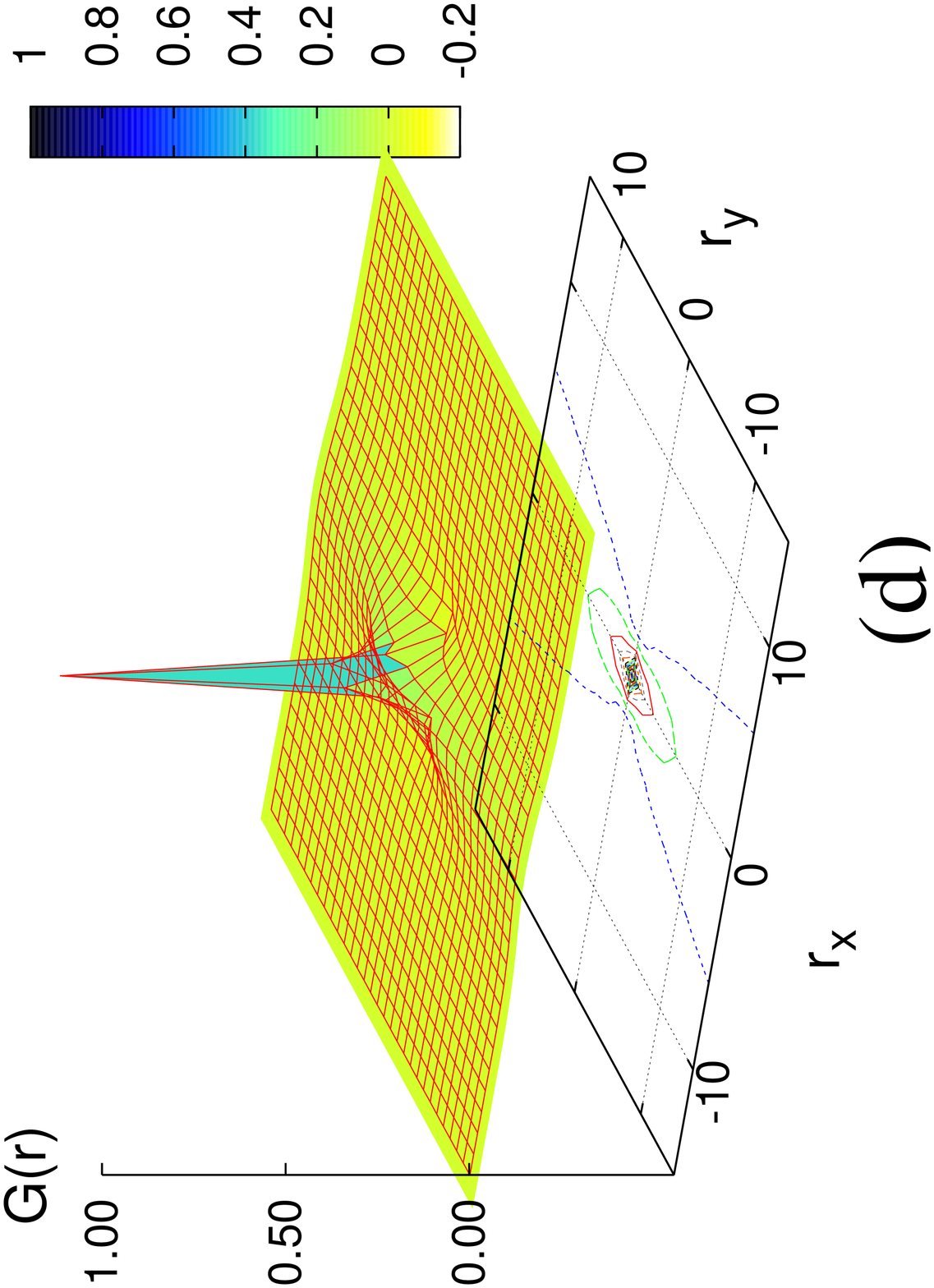, angle=-90}}
	  }
	  {\scalebox{0.19}
	     {\epsfig{file=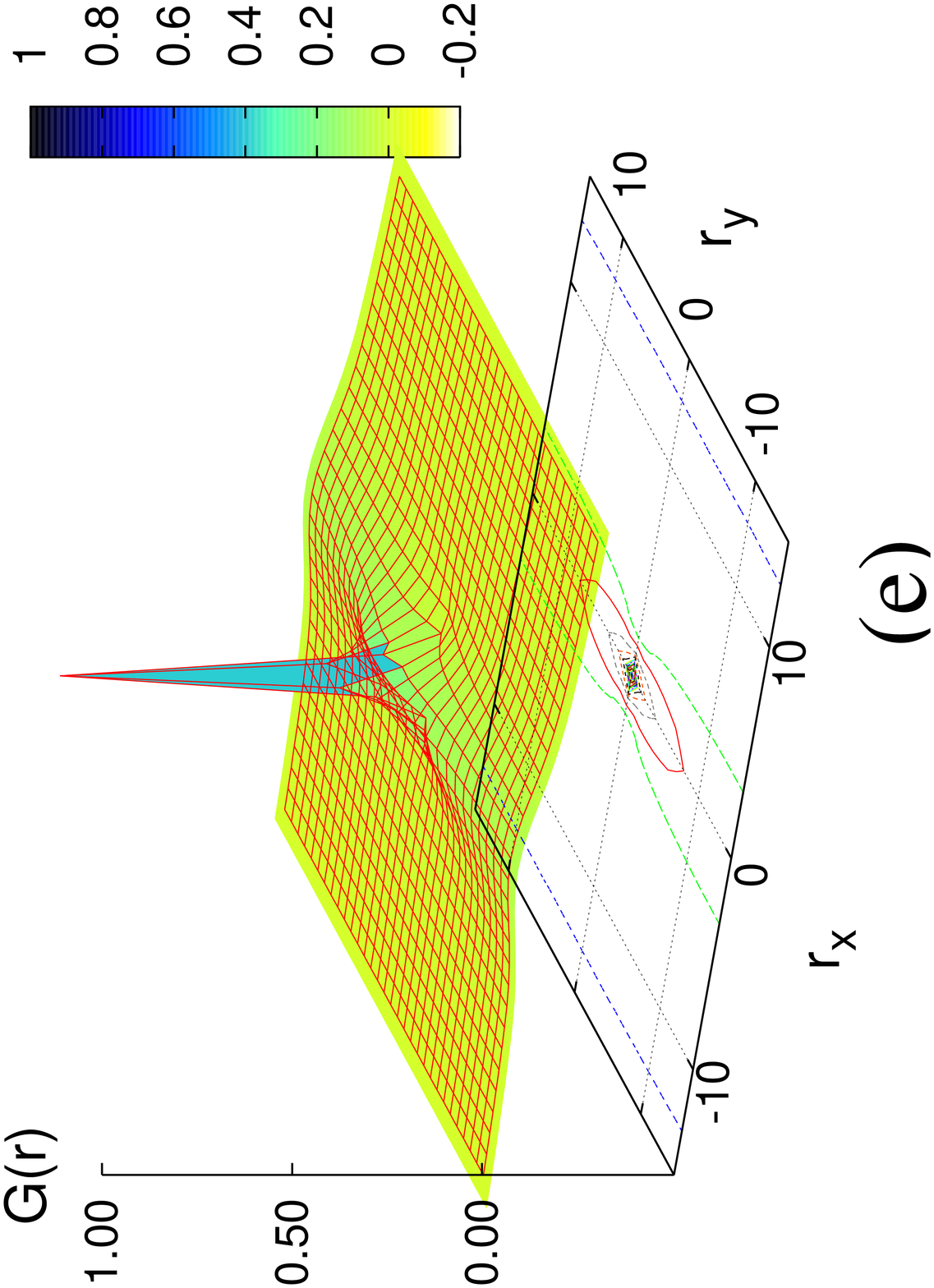, angle=-90}}
	  }
	  {\scalebox{0.19}
	     {\epsfig{file=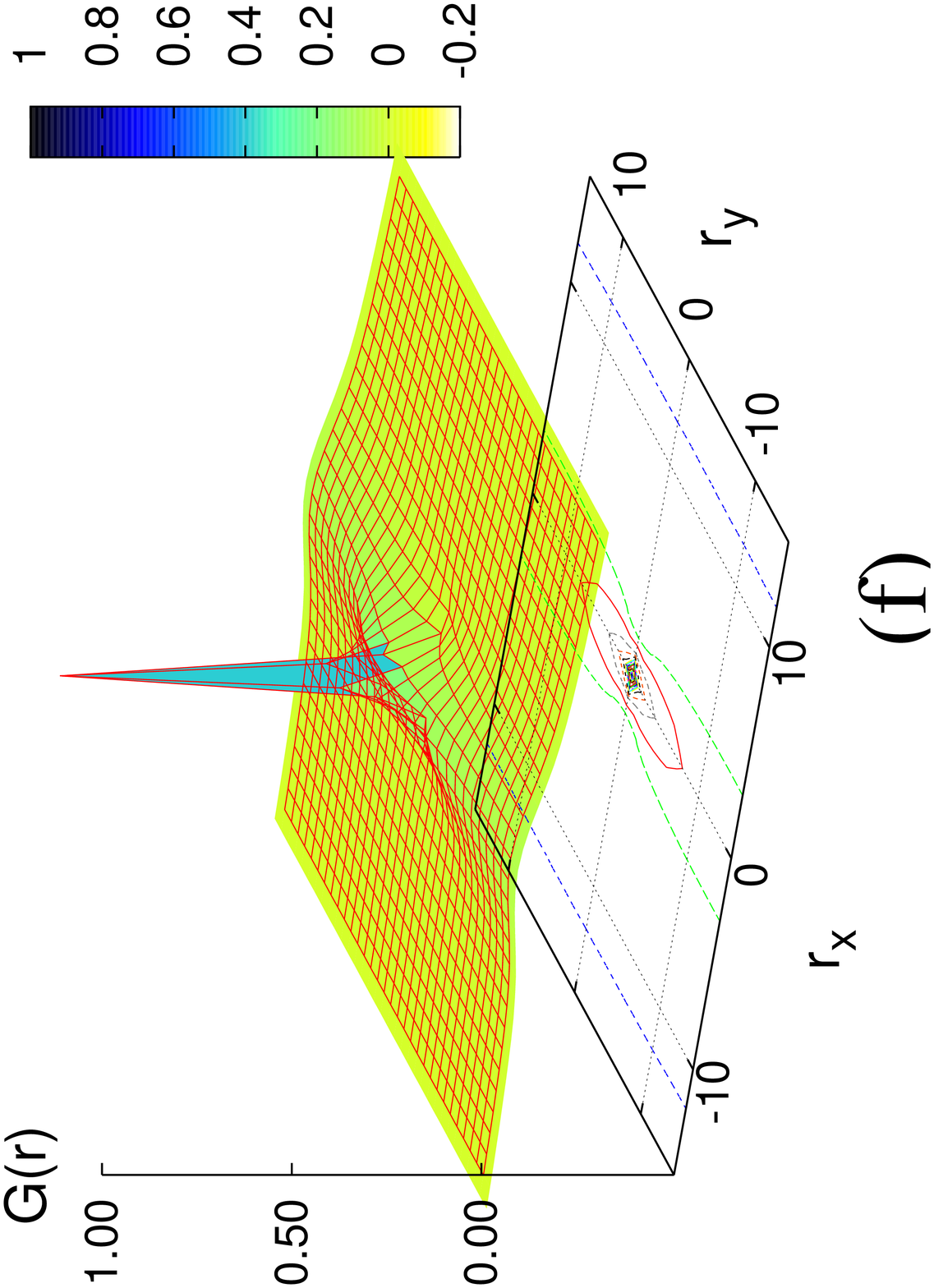, angle=-90}}
	  }
    }
\end{center}
\par
\vspace{-2mm}
\caption{The pair correlation function for the equilibrium system and its
driven counterpart on a $100\times 100$ lattice. 
The top row shows the Ising lattice gas at $T=2.47$ 
with Metropolis (a), heat bath (b), and Glauber (c) dynamics; the bottom
row shows the driven system at $T=3.60$ with Metropolis (d), heat bath (e), 
and Glauber (f) dynamics.}
\label{fig:surface}
\end{figure}

\begin{figure}[tbph]
\begin{center}
\mbox{ 
           {\scalebox{0.21}
	     {\epsfig{file=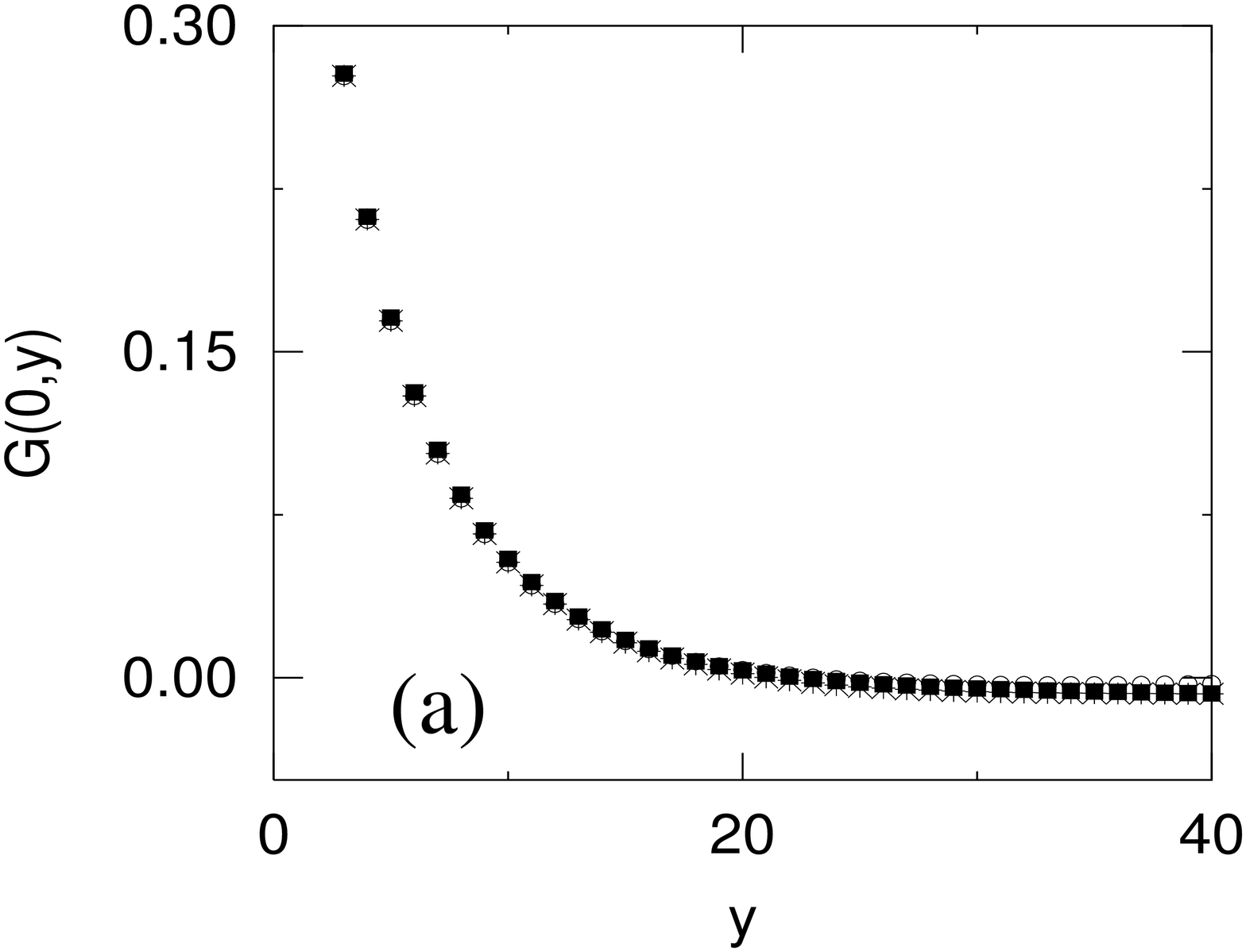, angle=-90}}
	  }\qquad\qquad
           {\scalebox{0.21}
	     {\epsfig{file=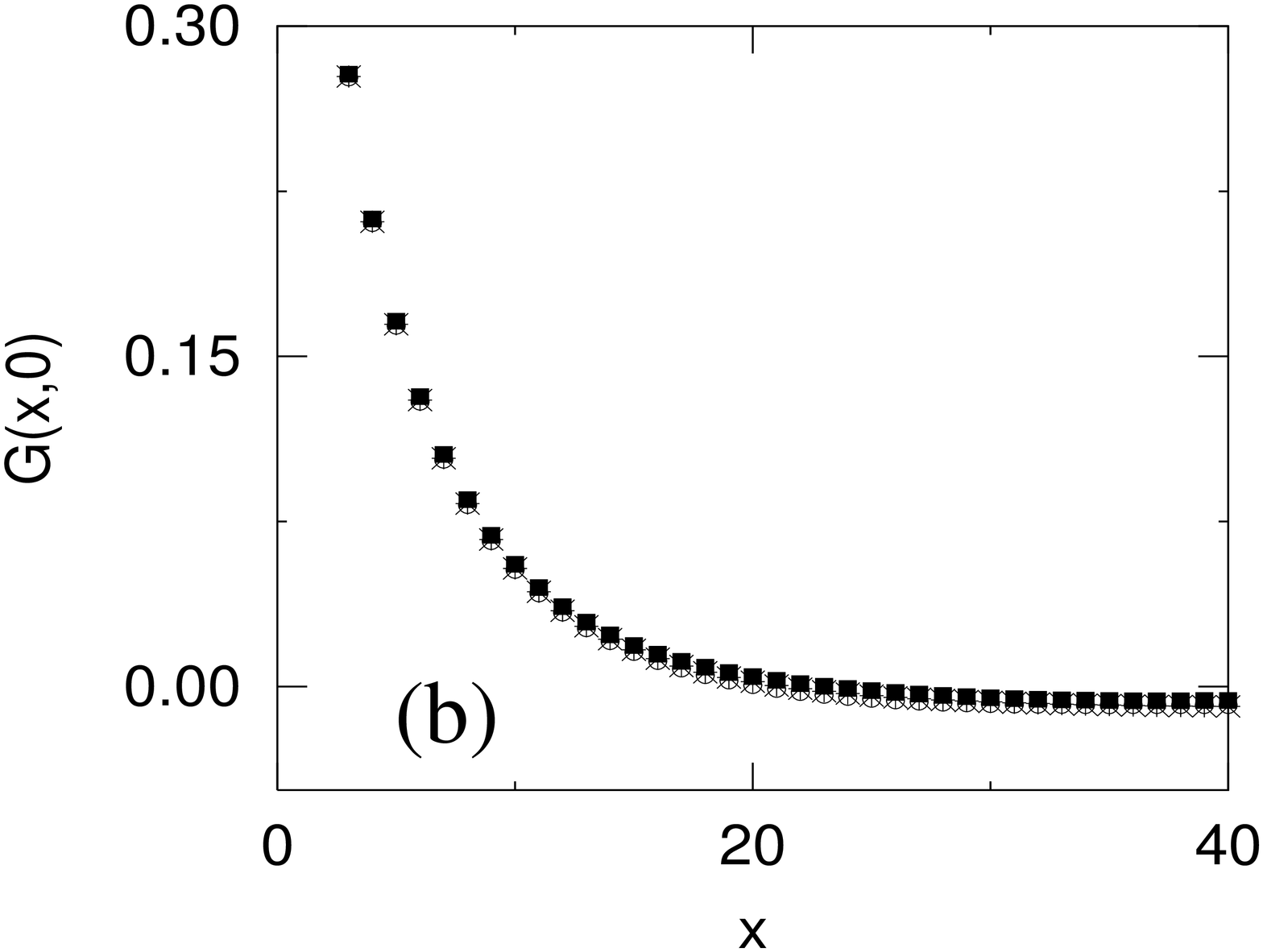, angle=-90}}
	  }
 } 
\mbox{ 
           {\scalebox{0.21}
	     {\epsfig{file=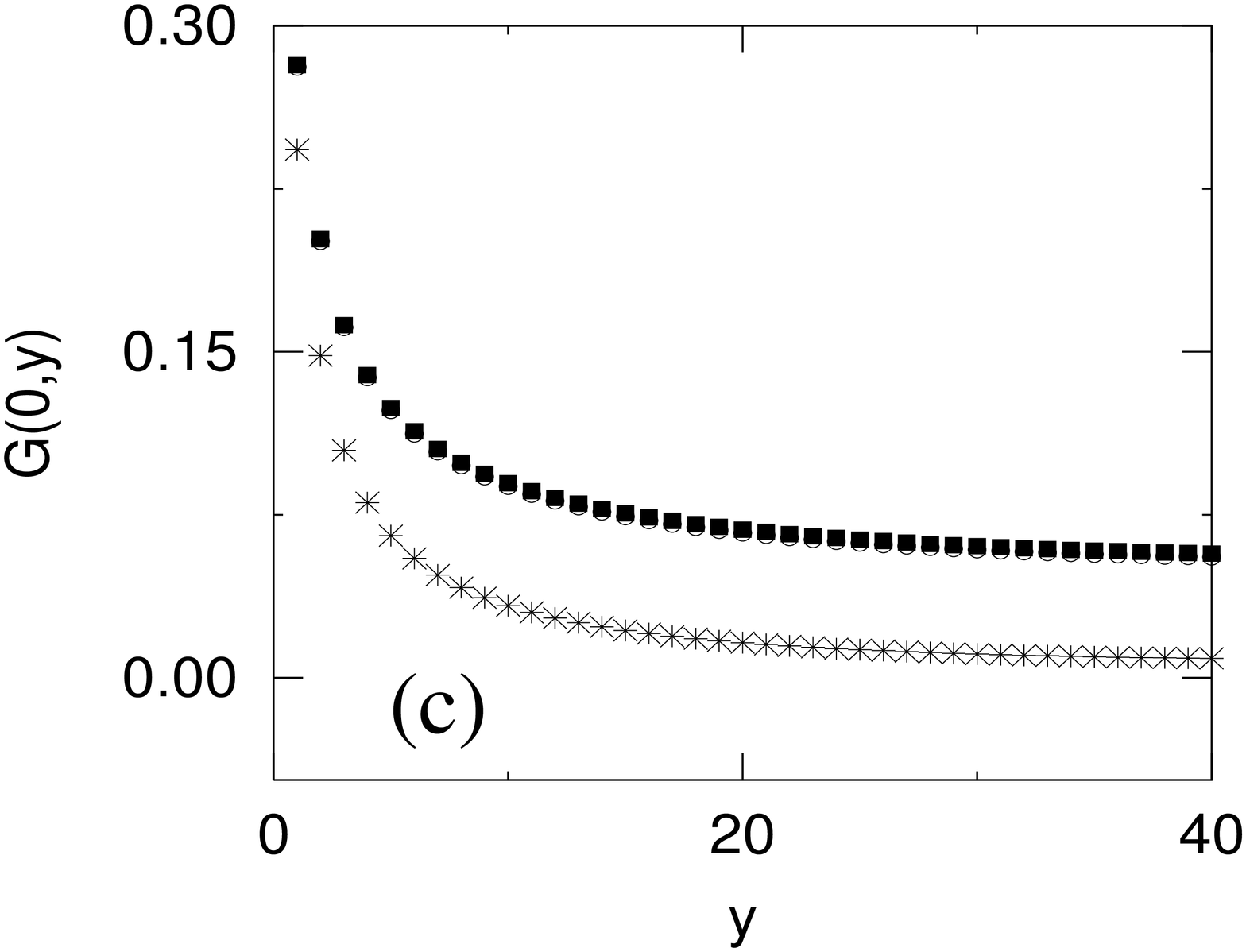, angle=-90}}
	  }\qquad\qquad
           {\scalebox{0.21}
	     {\epsfig{file=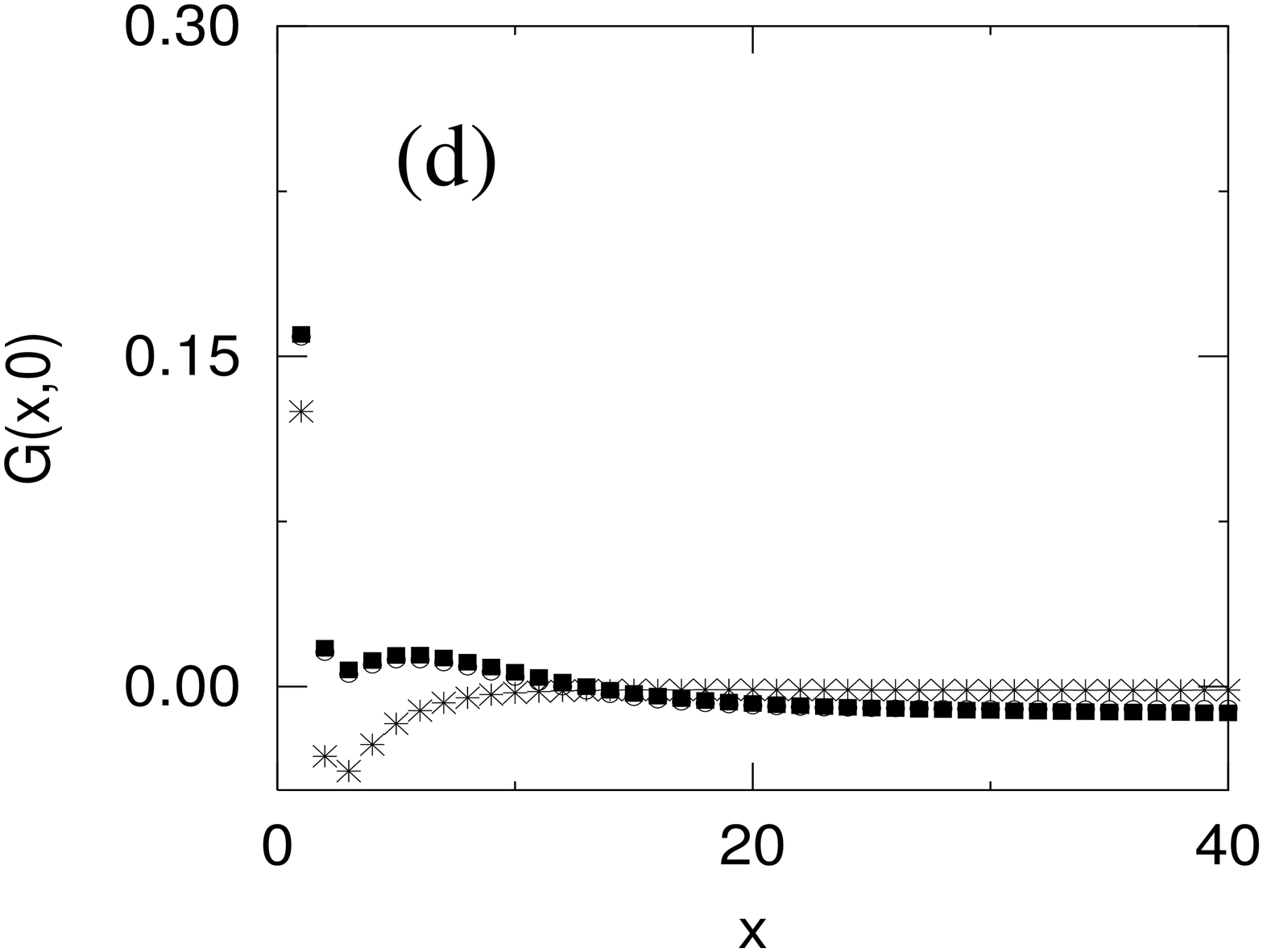, angle=-90}}
	  }
 }
\end{center}
\par
\vspace{-8mm}
\caption{Parallel and transverse two-point correlations for 
the equilibrium system at $T=2.47$ (a, b) and for the driven case
at $T=3.60$ (c, d) on a $100 \times 100$ lattice. 
Each plot shows data for three dynamics:
Metropolis (asterisks), heat bath (filled squares) and Glauber (open circles).
In (a, b) all data collapse, 
while in (c, d) only heat bath and Glauber data 
overlap. } 
\label{fig:ddsGy}
\end{figure}

\begin{figure}[tbph]
\begin{center}
\mbox{ 
	  {\scalebox{0.26}
	      {\epsfig{file=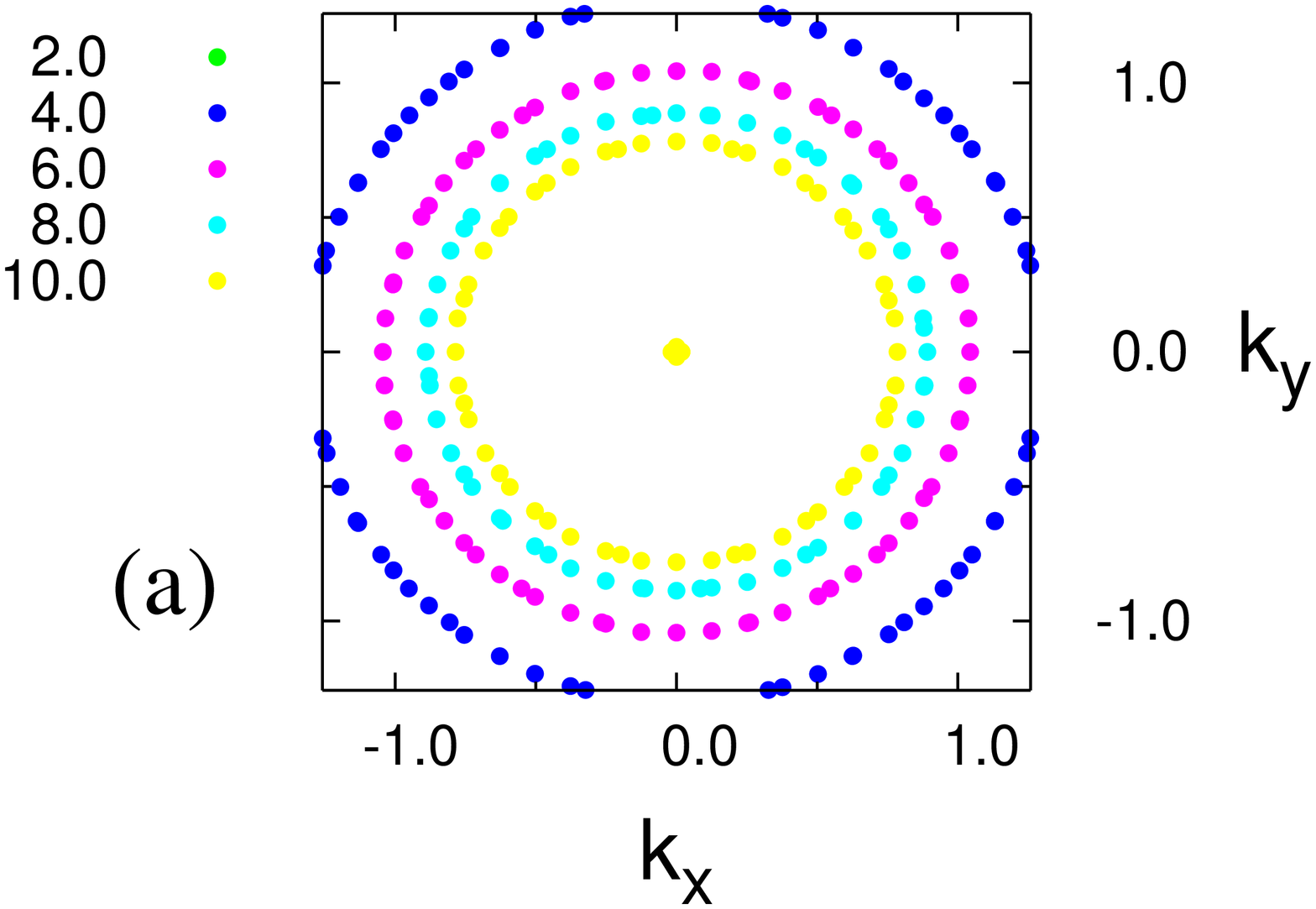, bbllx=150,bburx=580, bblly=0, bbury=620, angle=-90}}
	  }
	  {\scalebox{0.26}
	      {\epsfig{file=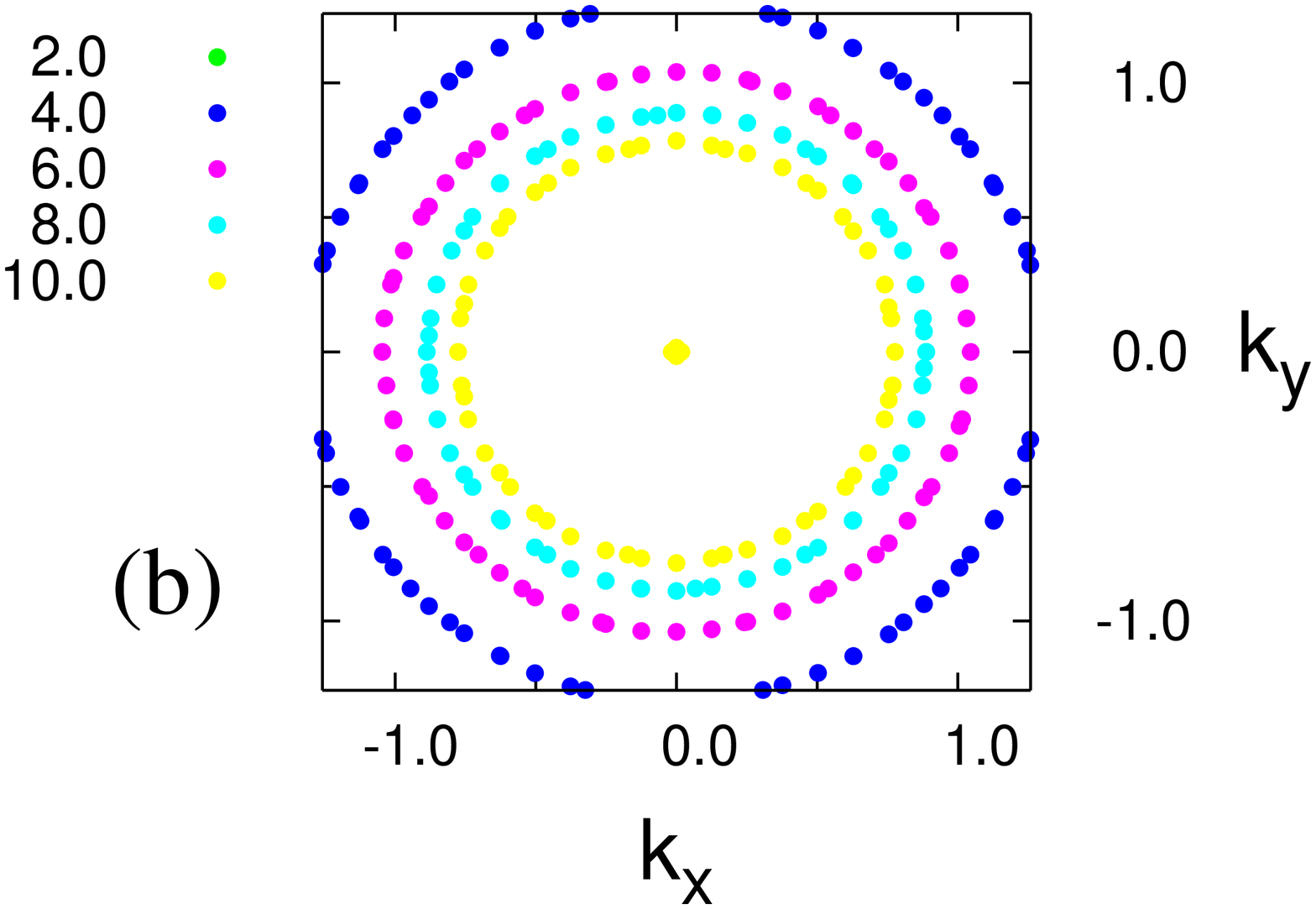, bbllx=150,bburx=580, bblly=0, bbury=620, angle=-90}}
	  }
	  {\scalebox{0.26}
	      {\epsfig{file=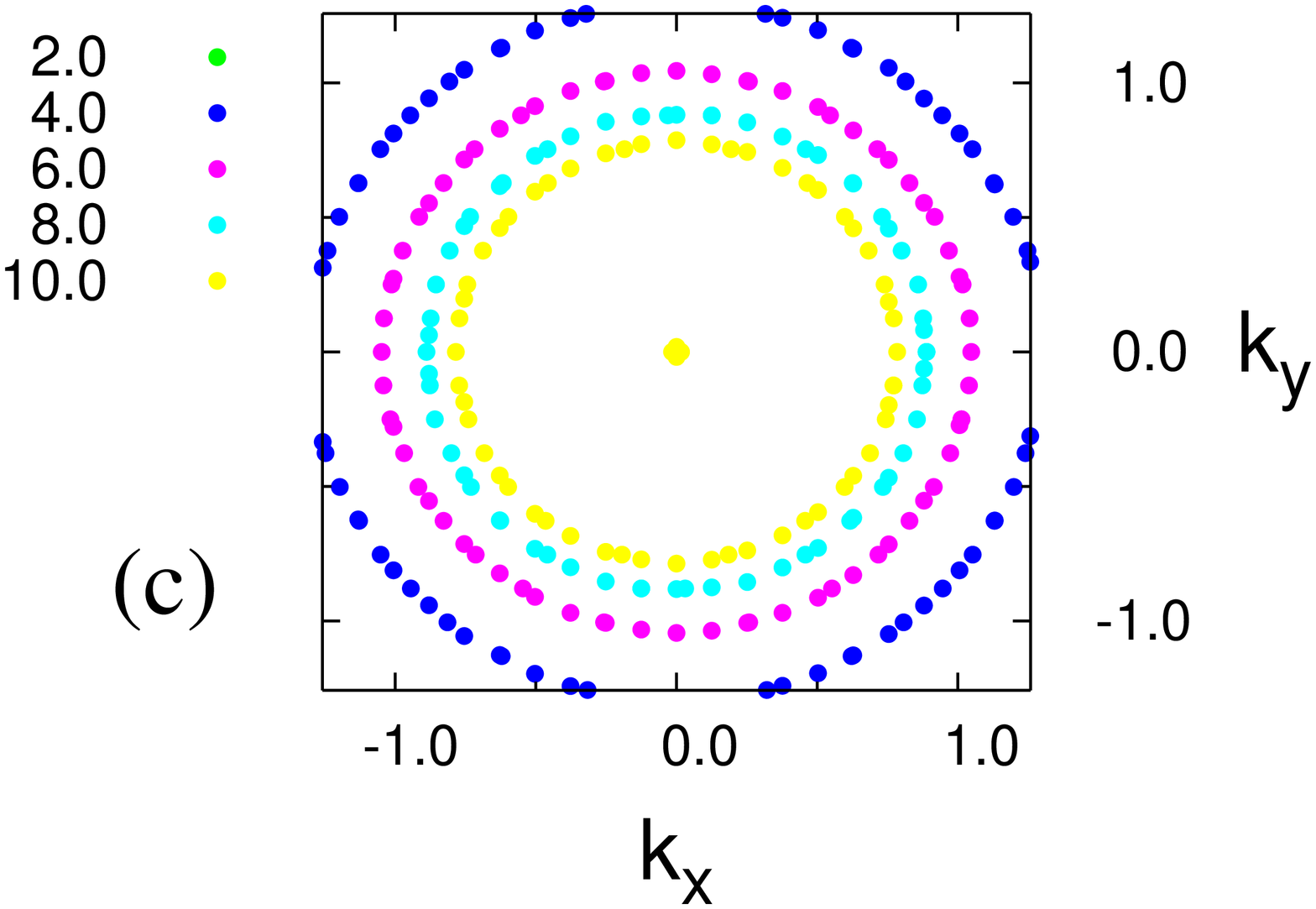, bbllx=150,bburx=580, bblly=0, bbury=620, angle=-90}}
	  }
    } 
\mbox{ 
	  {\scalebox{0.26}
	      {\epsfig{file=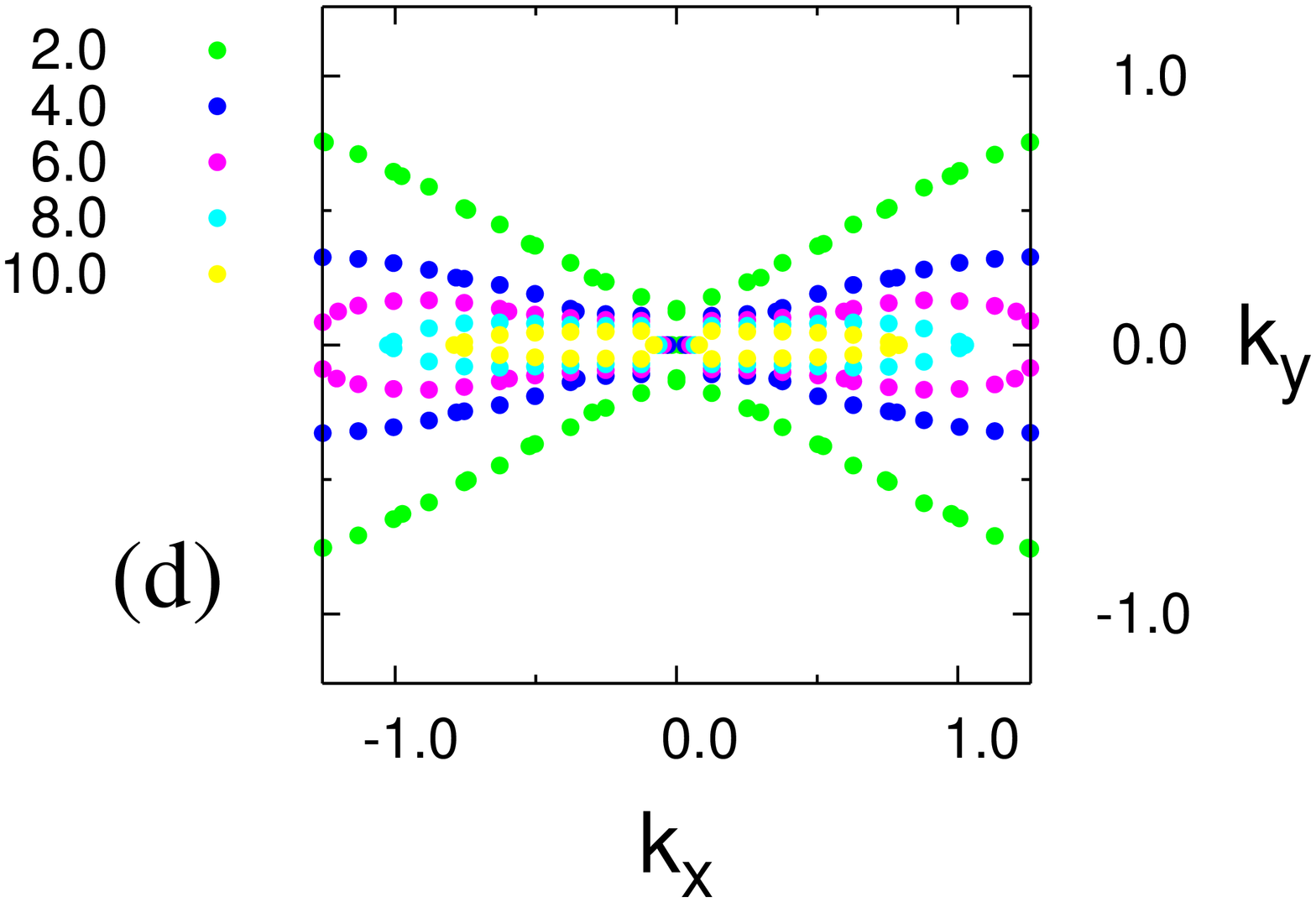, bbllx=150,bburx=580, bblly=0, bbury=620, angle=-90}}
	  }
	  {\scalebox{0.26}
	      {\epsfig{file=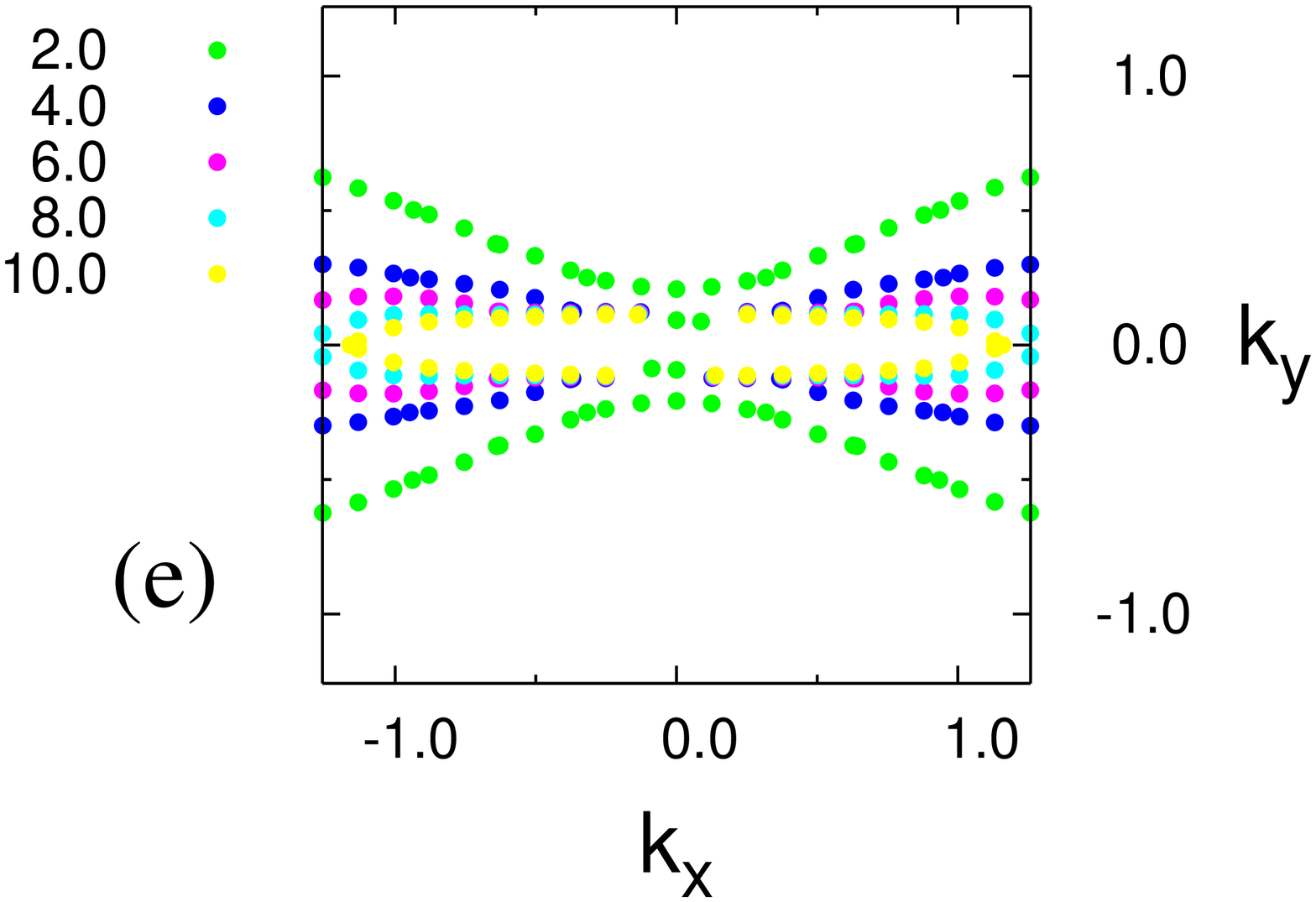, bbllx=150,bburx=580, bblly=0, bbury=620, angle=-90}}
	  }
	  {\scalebox{0.26}
	      {\epsfig{file=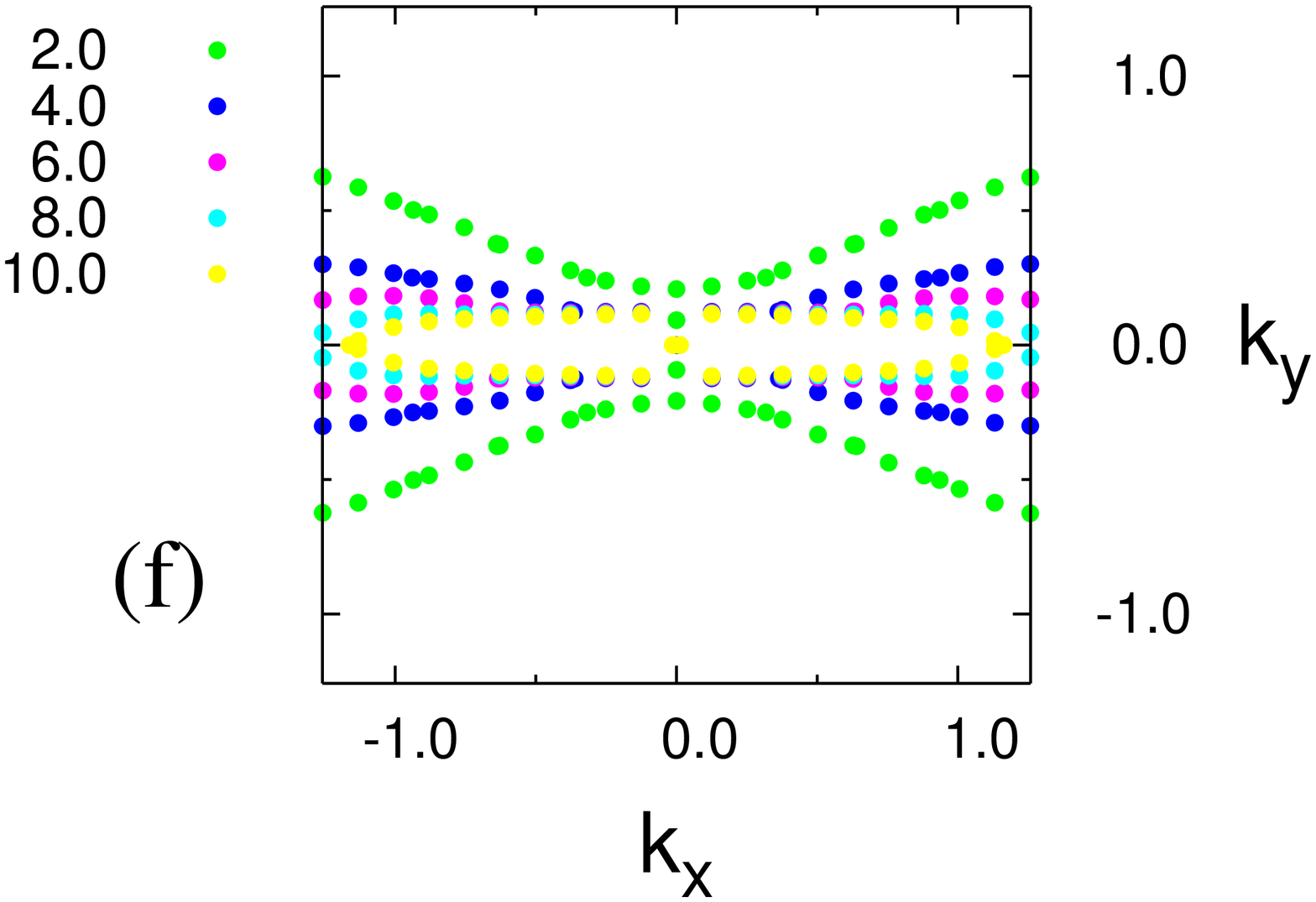, bbllx=150,bburx=580, bblly=0, bbury=620, angle=-90}}
	  }
    }
\end{center}
\par
\vspace{-4mm}
\caption{Structure factor contour plots for the equilibrium system and its
driven counterpart on a $100\times 100$ lattice. 
The top row shows the Ising lattice gas at $T=2.47$ 
with Metropolis (a), heat bath (b), and Glauber (c) dynamics; the bottom
row shows the driven system at $T=3.60$ with Metropolis (d), heat bath (e), 
and Glauber (f) dynamics.}
\label{fig:contour}
\end{figure}

\begin{figure}[tbph]
\begin{center}
\mbox{ 
           {\scalebox{0.21}
	     {\epsfig{file=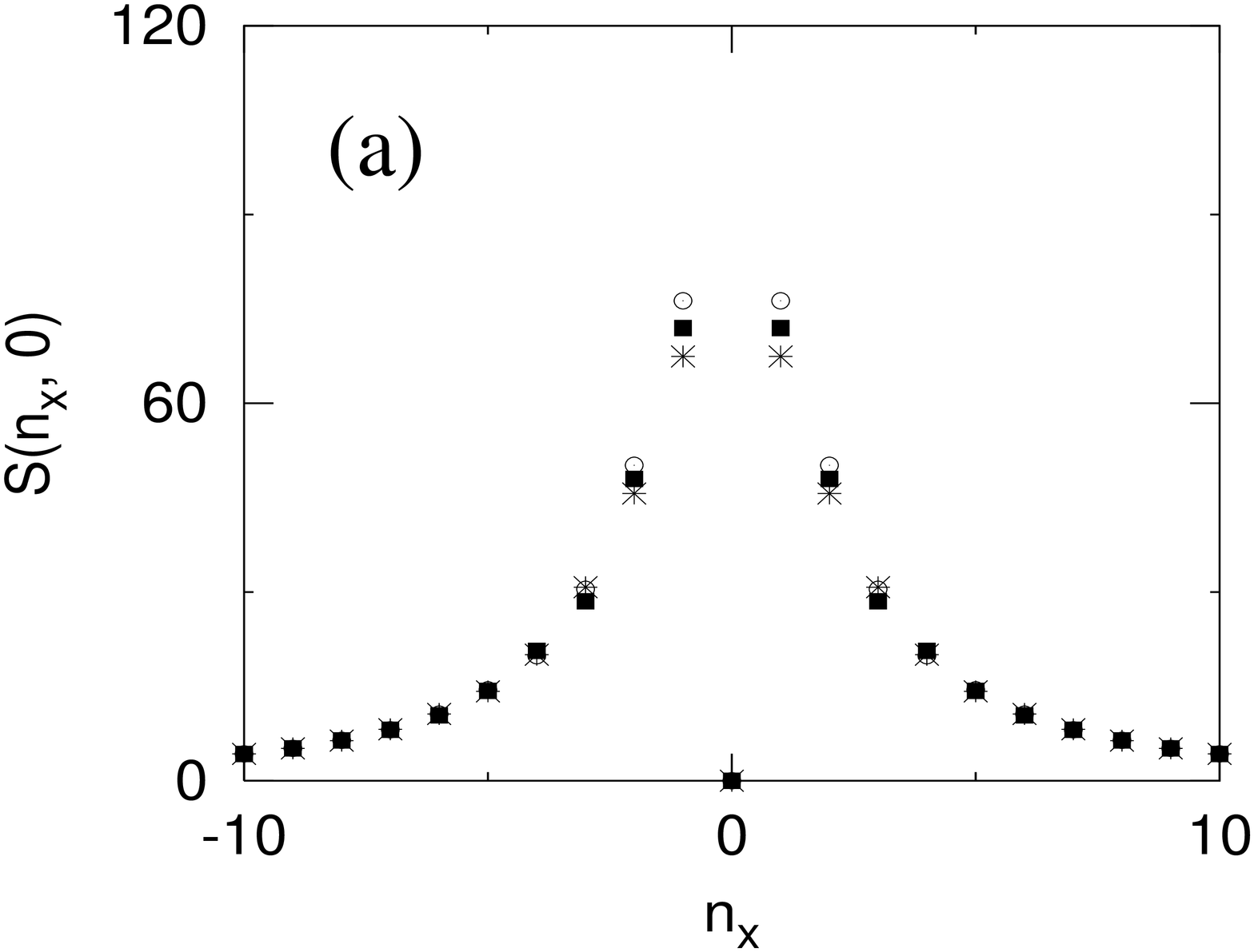, angle=-90}}
	  }\qquad\qquad
           {\scalebox{0.21}
	     {\epsfig{file=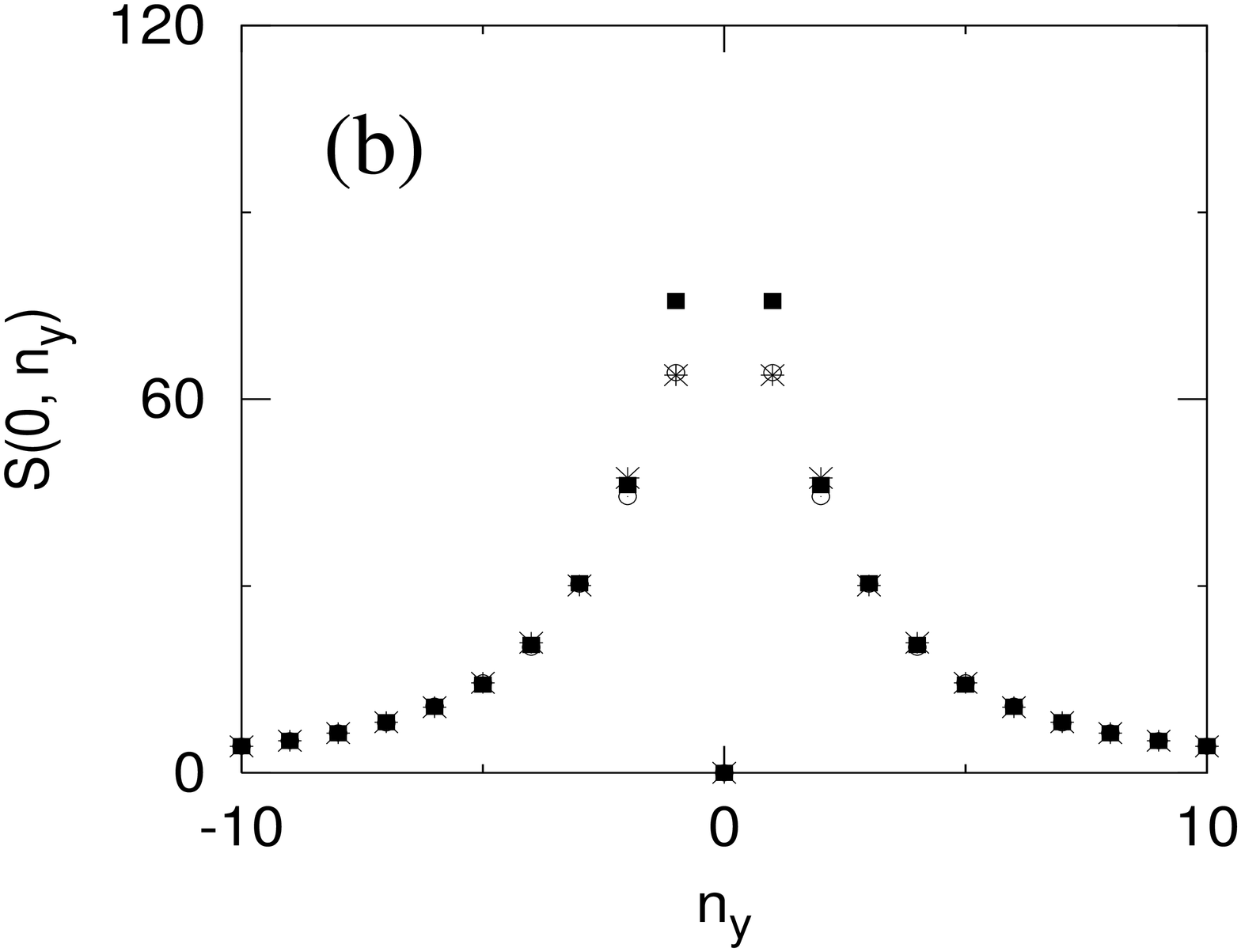, angle=-90}}
	  }
 } 
\mbox{ 
           {\scalebox{0.21}
	     {\epsfig{file=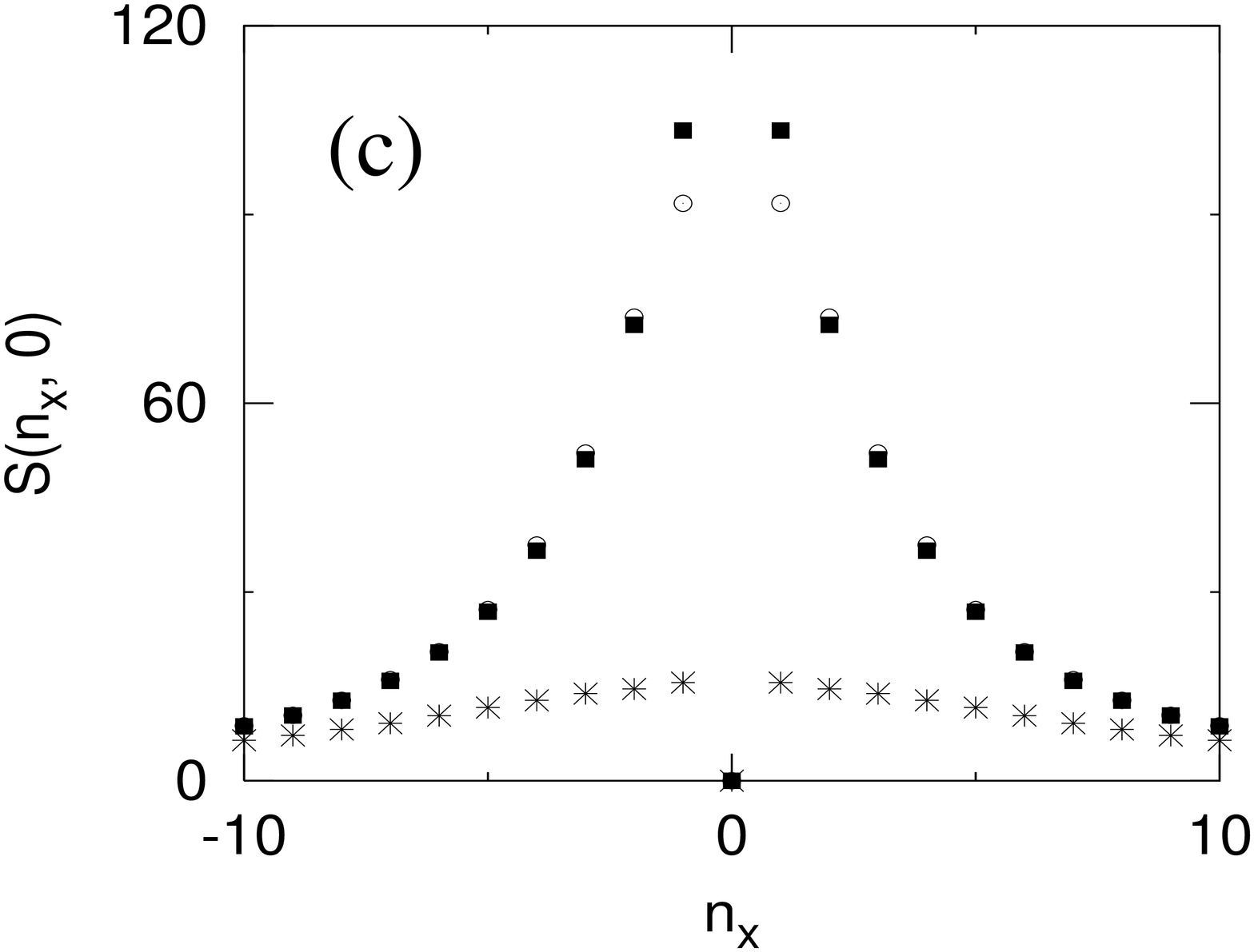, angle=-90}}
	  }\qquad\qquad
           {\scalebox{0.21}
	     {\epsfig{file=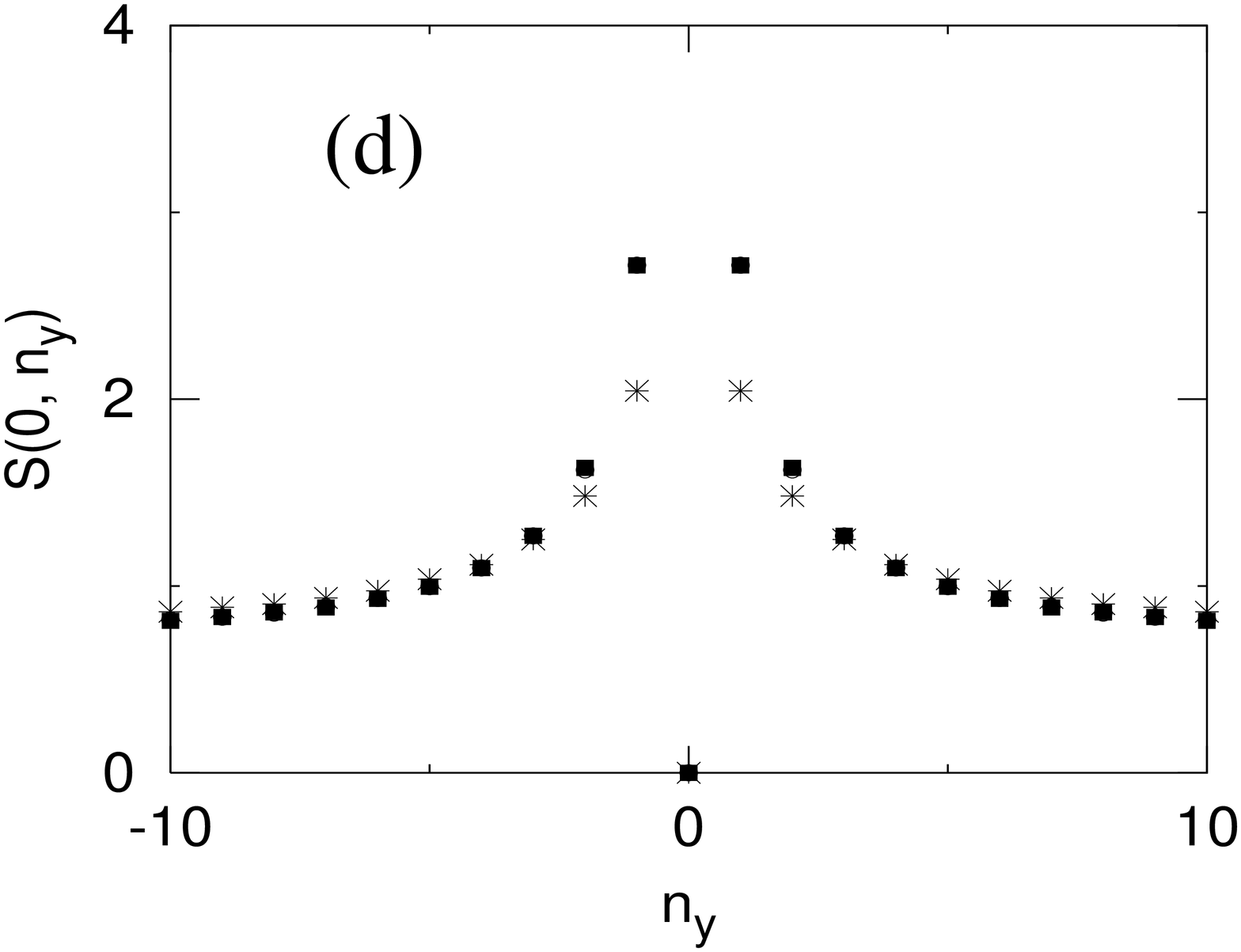, angle=-90}}
	  }
 }
\end{center}
\par
\vspace{-5mm}
\caption{Parallel and transverse structure factors for  
the equilibrium system at $T=2.47$ (a, b) and for the driven case
at $T=3.60$ (c, d), on a $100 \times 100$ lattice. 
Each plot shows data for three dynamics:
Metropolis (asterisks), heat bath (filled squares) and Glauber (open circles).
Within error bars (not shown), the data effectively collapse in (a, b), 
while only heat bath and Glauber data overlap in (c, d). 
Note the different scale in (d).} 
\label{fig:ddsSx}
\end{figure}

\end{widetext}

\subsection{Typical configurations, two-point correlations and structure
factors.}

We begin our discussion by showing a few typical configurations of the
driven system on a $48\times 432$ lattice. Figs.~2a and b are obtained for
the equilibrium case, just below and above criticality. The preference for
horizontal interfaces is clearly seen in Fig.~2a. The remaining configurations
(Fig.~2c-h) all show the driven system, for heat bath and
Metropolis dynamics, at three different temperatures. At the lowest
temperature $T_{1}=2.90$ (Figs.~2c, d), the driven system is ordered for
both dynamics. In stark contrast to the equilibrium case, the interfaces
between high- and low-density regions are parallel to $\mathcal{E}$ and
therefore clearly \emph{not} dominated by energetics. At a slightly higher
temperature, $T_{2}=3.30$ (Figs.~2e, f), we observe the first glaring
discrepancy between the two dynamics: the configuration generated by the
heat bath algorithm is still ordered while the Metropolis configuration is
already disordered! Eventually, at $T_{3}=3.70$, both algorithms generate
disordered configurations. Clearly, the two algorithms lead to \emph{%
different} critical temperatures, with $T_{c}^{M}<$ $T_{c}^{H}$. A rough
estimate based on our data \cite{WK-unpub} results in $T_{c}^{H}=3.55\pm
0.05 $ and $T_{c}^{M}=3.15\pm 0.05$. More precise estimates \cite{Italy} are
available for Metropolis rates only: $T_{c}^{M}=3.\,\allowbreak 198\,01(19)$.

To probe this apparent discrepancy between Metropolis and heat bath rates
further, and to explore the position of Glauber rates in this triad, we turn
to a more detailed analysis. In Fig.~3, we show surface plots of $G\left( 
\mathbf{r}\right) $ for the Ising lattice gas (top row, Figs.~3a-c) and the
driven system (bottom row, Figs.~3d-f). The three columns correspond to the
three different dynamics: Metropolis (Figs.~3a, d), heat bath (Figs.~3b, e),
and Glauber (Figs.~3c, f). Fig.~4 shows selected projections of $G\left( 
\mathbf{r}\right) $, namely $G(0,y)$ and $G(x,0)$, for equilibrium (Fig.~4a,
b) and with infinite drive (Fig.~4c, d). As dictated by detailed balance, the
correlation functions for the \emph{equilibrium system} are independent of
dynamics: there are no discernable differences between Figs.~3a-c, and the
data in Figs.~4a and b collapse within statistical error bars (less than 
$0.01$ in absolute units). The chosen temperature, $T=2.47$, is close enough 
to Ising criticality so that lattice anisotropies are irrelevant: 
$G\left( \mathbf{r}\right) $ is isotropic, with circular contours 
centered on the origin. The small negative values observed at large 
distances are a consequence of the sum rule.

This simple picture becomes considerably more complex when we turn to the 
\emph{driven} system (lower row of Fig.~3 and Figs.~4c, d). The chosen
temperature, $T=3.60$, is very close to our estimate for the critical
temperature of heat bath and Glauber rates, $T_{c}^{H}\simeq T_{c}^{G}$ $%
\simeq 3.55$ and about $15\%$ above $T_{c}^{M}$. We immediately note the
strong anisotropy induced by the drive. Further, there are noticeable
differences between Metropolis rates on one hand, and heat bath and Glauber
dynamics on the other. These are most easily observed in Figs.~4c and d. For
Metropolis rates, $G^{M}(0,y)$ is positive and decreases monotonically
throughout (Fig.~4c), while $G^{M}(x,0)$ drops rapidly below zero, displays
a minimum and then recovers and approaches zero from below. These features
have been noted before \cite{KLS}, and are directly related to the breaking
of detailed balance \cite{SZ,corr'ns-anal,discont}. The data for Glauber and
heat bath dynamics, while practically indistinguishable from one another,
differ visibly from the Metropolis ones.\ Considering correlations measured
along the field direction first, we observe $G^{H}(0,y)\simeq
G^{G}(0,y)>G^{M}(0,y)$ for all $y$. In other words, Metropolis rates
generate weaker correlations, consistent with the lower $T_{c}^{M}$. Heat
bath and\ Glauber rates produce roughly the same correlations; moreover,
these show clear signatures of being very close to criticality, evidenced by
the distinctly positive value at the largest $y$ shown: $G^{H}(0,40)\simeq
G^{G}(0,40)\simeq 0.07$. Highly correlated domains in the driven system are
needle-shaped, with the needle pointing along the field, and this small, yet
nonzero value indicates that some of these domains are long enough to span
half the system. These precursors of ordering become even more obvious when
we turn to correlations \emph{transverse} to the field: The secondary maximum
in Fig.~4d indicates a tendency towards forming thin stripes for heat bath
and Glauber rates. 

The structure factors bear out this picture. Again, the independence from
the rates, and the isotropy near criticality is clearly displayed by the
contour plots for the equilibrium system, shown in Figs.~5a-c, and by the
projections shown in Figs.~6a, b. In the driven case (Figs.~5d-f and 6c, d),
the presence of strong anisotropy is apparent, and the well-known
discontinuity singularity at the origin \cite{discont} is observed easily: $%
\lim_{k_{x}\rightarrow 0}S(k_{x},0)\neq \lim_{k_{y}\rightarrow 0}S(0,k_{y})$%
. While these broad features characterize all three dynamics, the absolute
values of the structure factors differ slightly from one another: $%
S^{M}(k_{x},k_{y})$ \ is generally smaller than either $S^{H}(k_{x},k_{y})$
or $S^{G}(k_{x},k_{y})$. Moreover, the distance from criticality can be
measured through the discontinuity\emph{\ ratio},

\[
\mathcal{S}\equiv \frac{\lim_{k_{x}\rightarrow 0}S(k_{x},0)}{%
\lim_{k_{y}\rightarrow 0}S(0,k_{y})} 
\]%
which diverges as $T\rightarrow T_{c}$ \cite{discont}. Our data result in 
$\mathcal{S}^{M}\simeq 7.5$, while 
$\mathcal{S}^{H}\simeq \mathcal{S}^{G}\simeq 34>\mathcal{S}^{M}$.
Our findings confirm, once again, that the heat bath and Glauber data are
effectively much closer to criticality than those for Metropolis rates.

\begin{figure}[tbph]
\begin{center}
\mbox{ 
	  {\scalebox{0.25}
	     {\epsfig{file=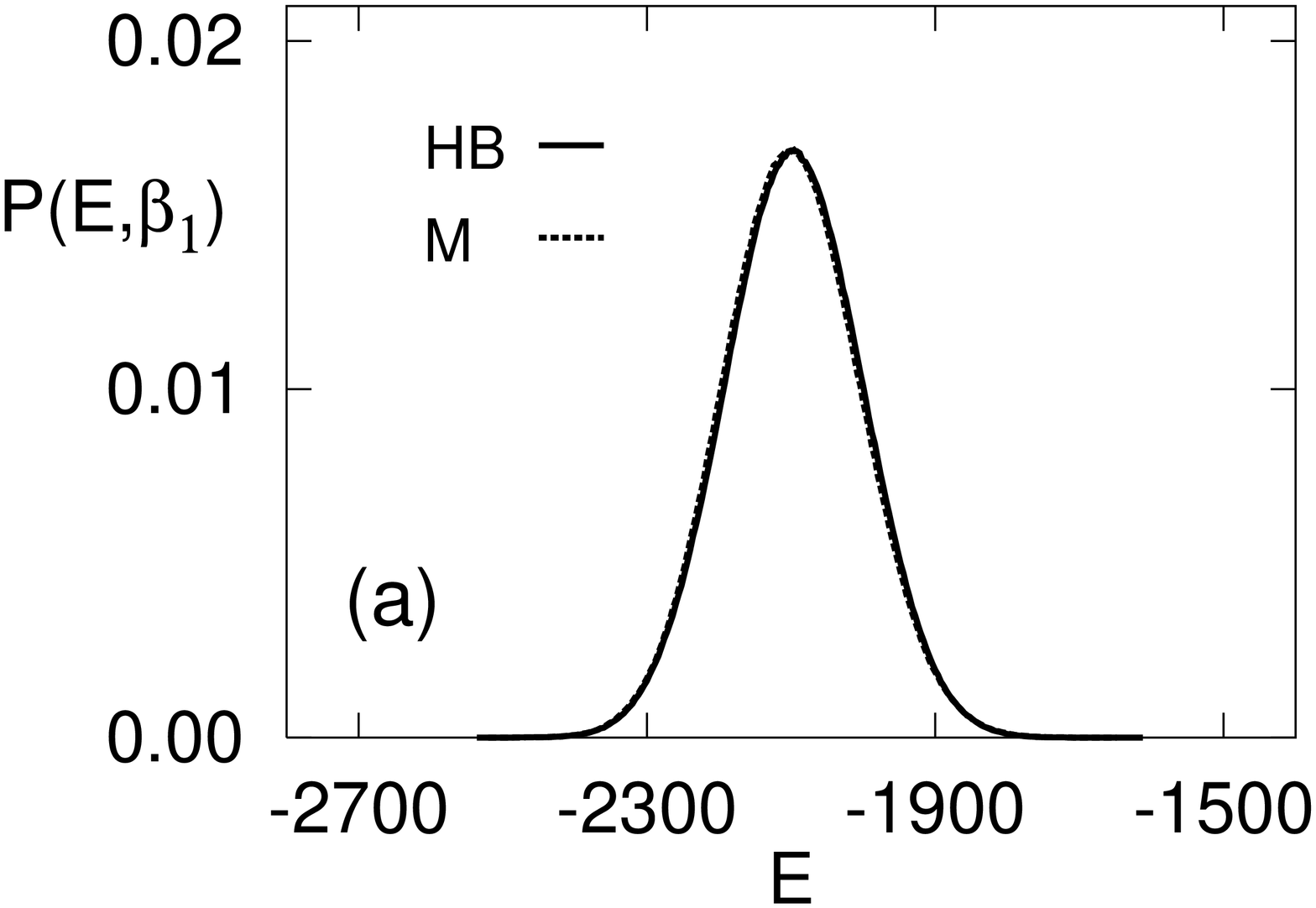, angle=-90}}
	  }
    } 
\mbox{ 
          {\scalebox{0.25}
	     {\epsfig{file=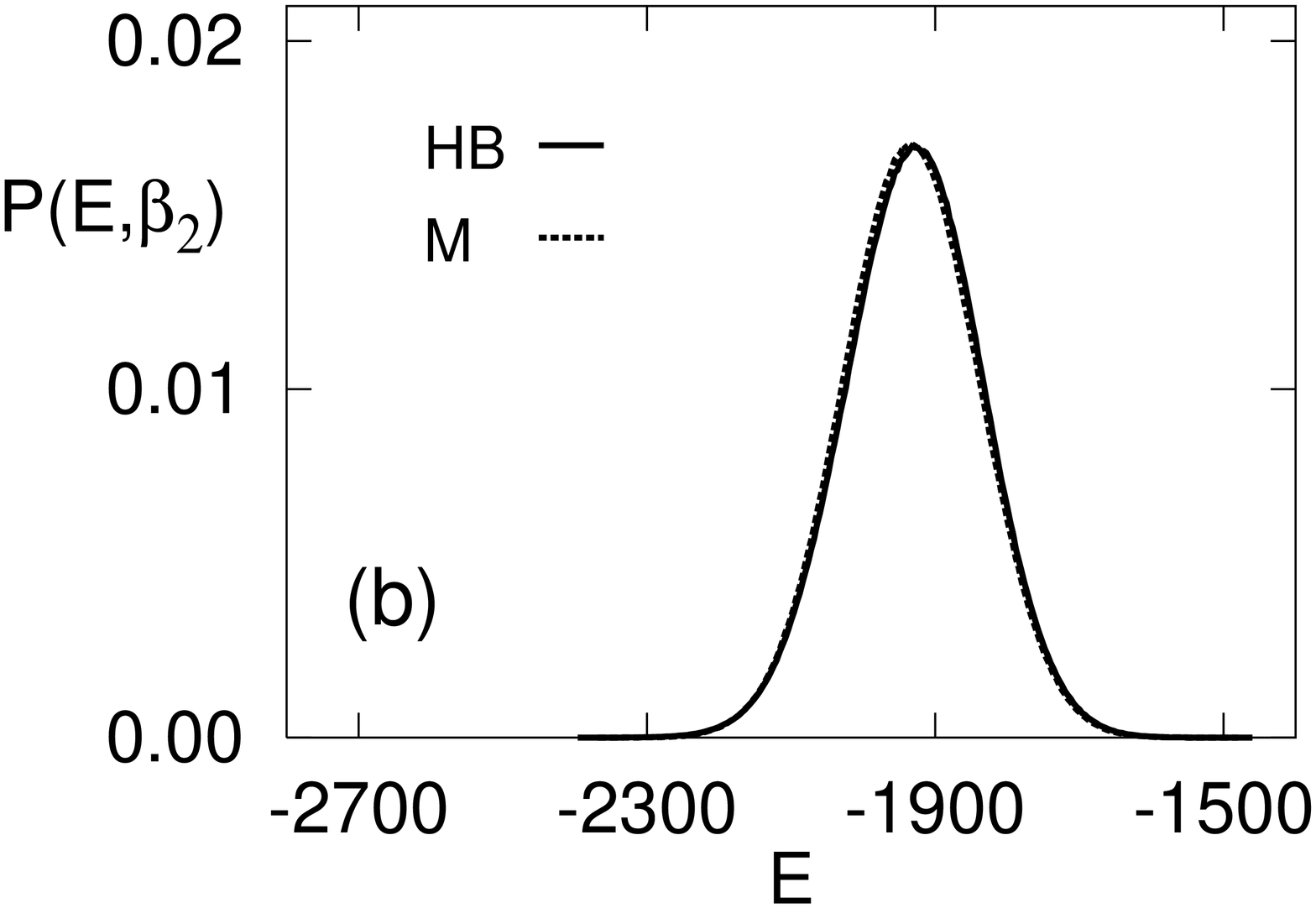, angle=-90}}
	  }
    } 
\mbox{ 
          {\scalebox{0.245}
	   {\epsfig{file=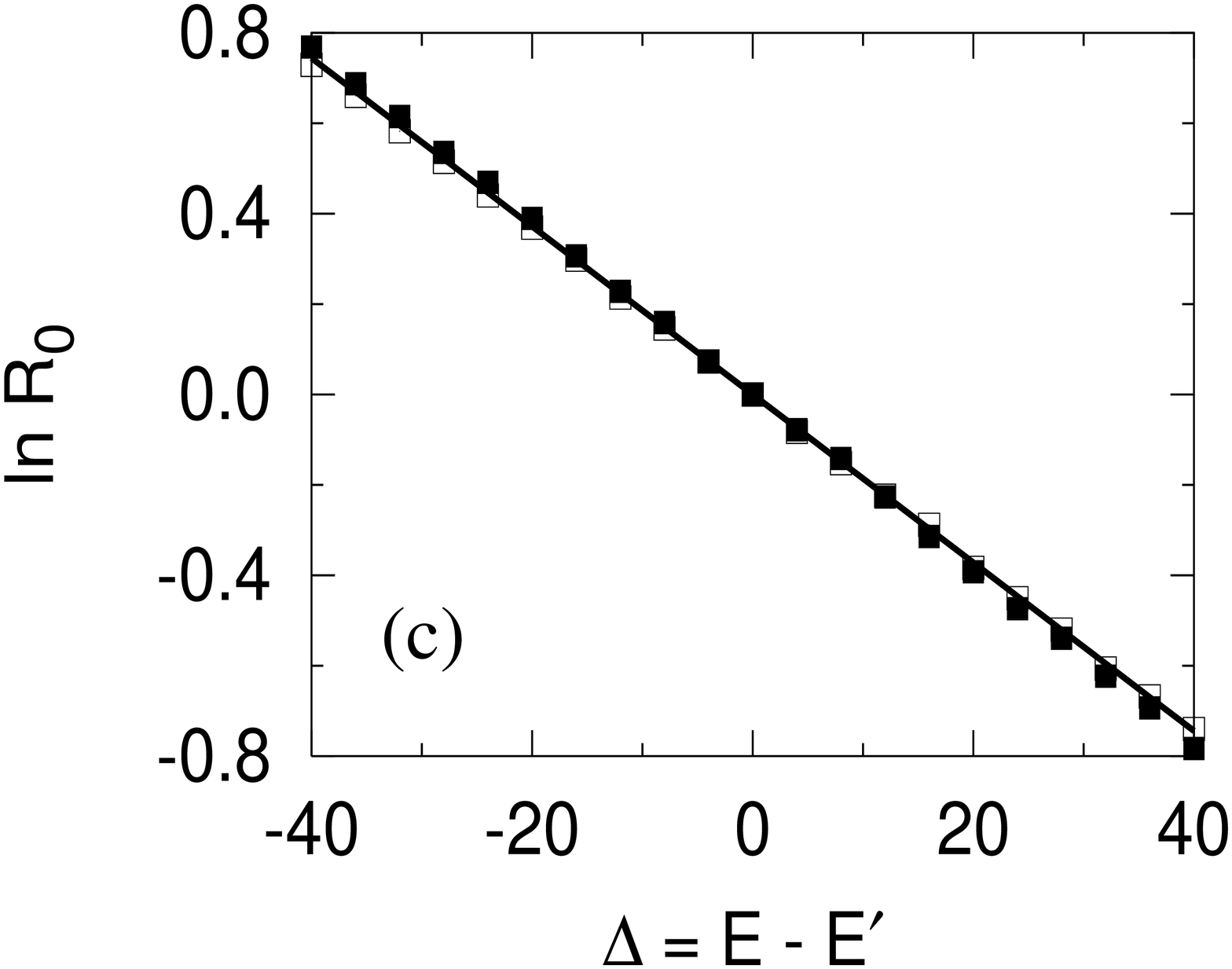, bbllx=20,bburx=630, bblly=100, bbury=750, angle=-90}}
	  }
    }
\end{center}
\par
\vspace{-6mm}
\caption{Normalized histograms for the equilibrium system using 
heat bath (solid line) and Metropolis (dotted line) rates, at 
$\beta_{1}=1/2.269$ (a) and $\beta_{2}=1/2.369$ (b).
In (c), we show $\ln R_{0}$ vs $E - E^{\prime}$: 
Data are shown as open (Metropolis) and filled (heat bath) squares; 
the solid line is the expected behavior,  
$-(\beta_{1}-\beta _{2})(E-E^{\prime })$. } 
\label{fig:histocomp1}
\end{figure}

\begin{figure}[tbph]
\begin{center}
\mbox{ 
	  {\scalebox{0.25}
	     {\epsfig{file=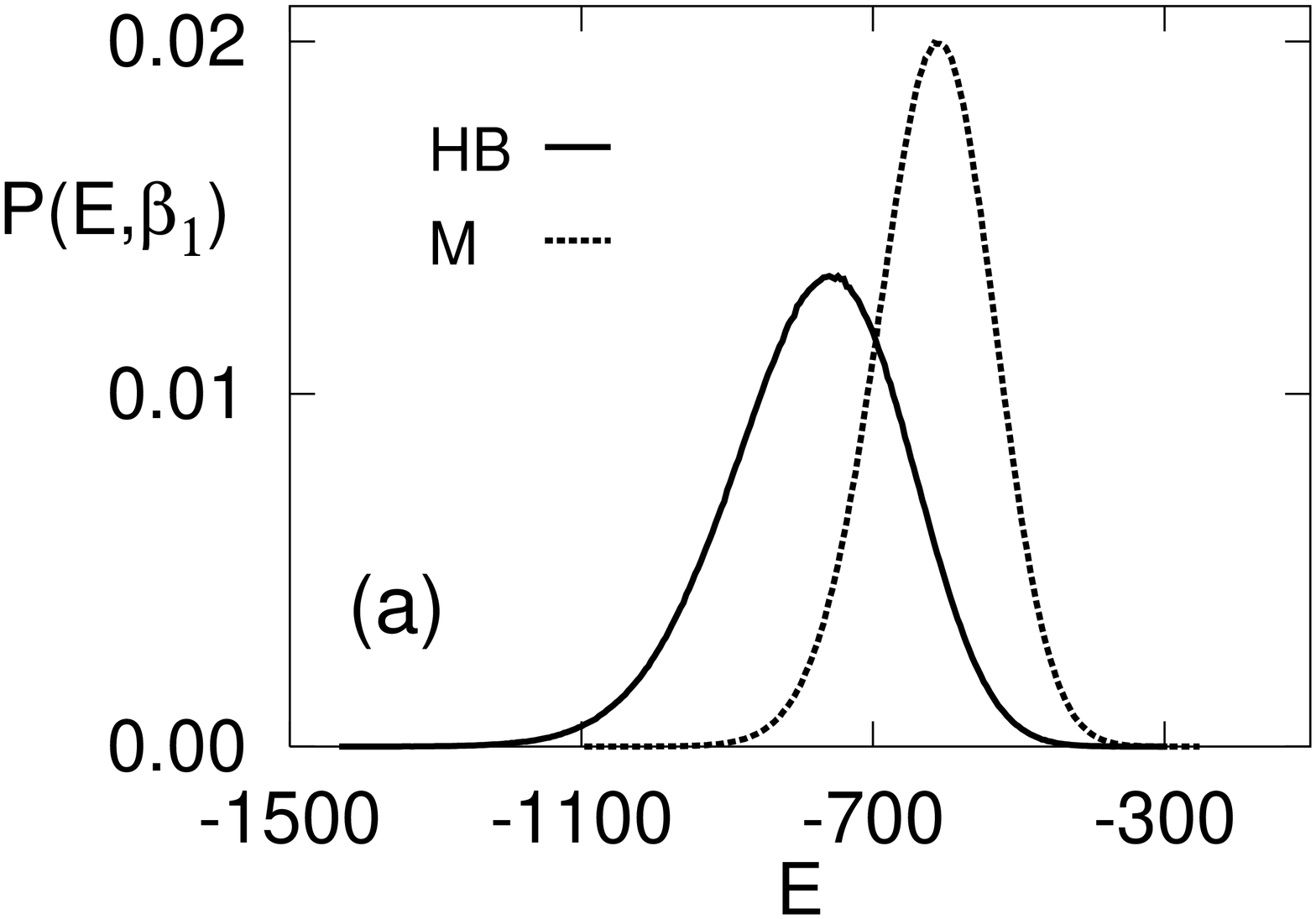, angle=-90}}
	  }
    } 
\mbox{ 
          {\scalebox{0.25}
	     {\epsfig{file=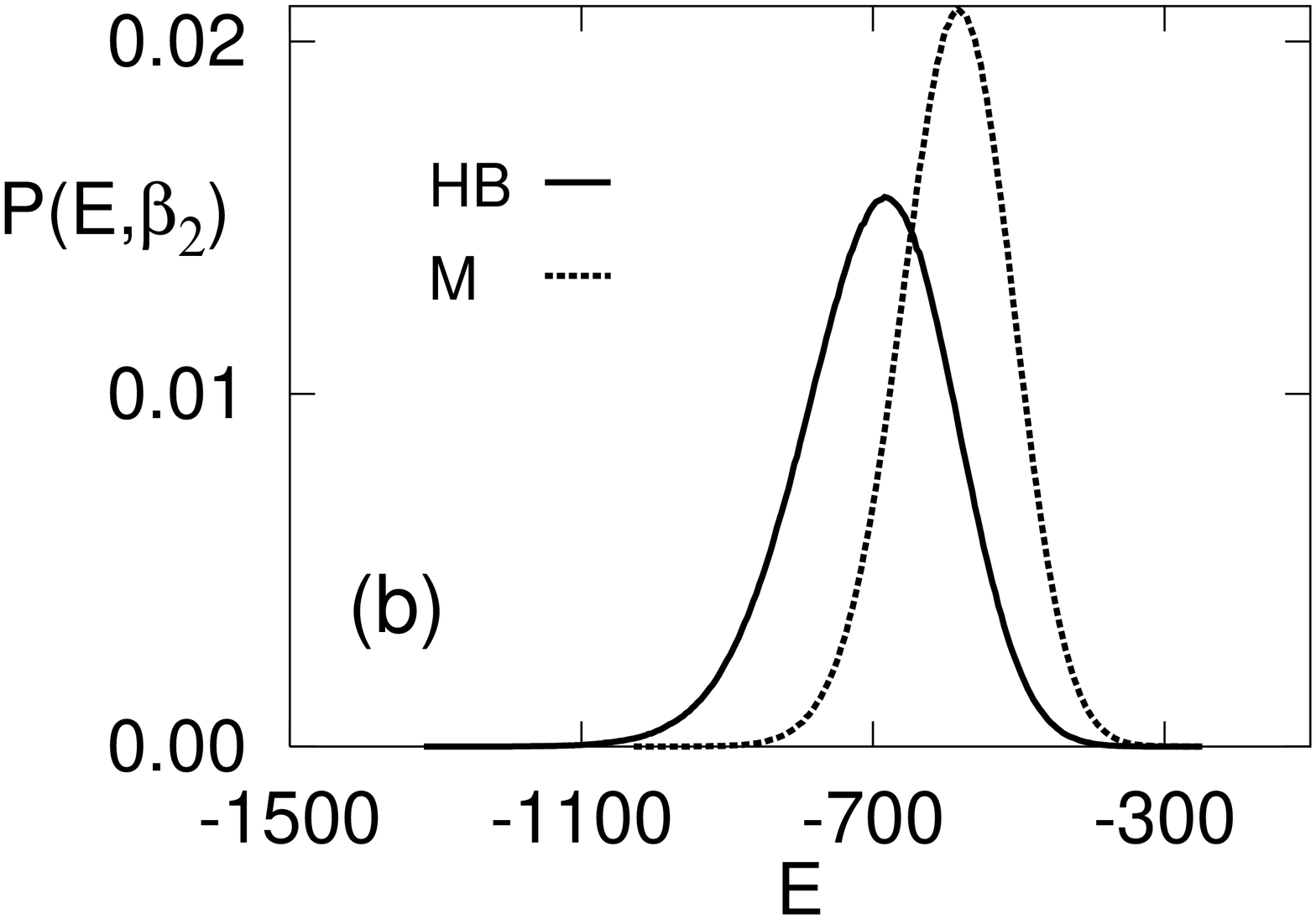, angle=-90}}
	  }
   } 
\mbox{ 
          {\scalebox{0.245}
	   {\epsfig{file=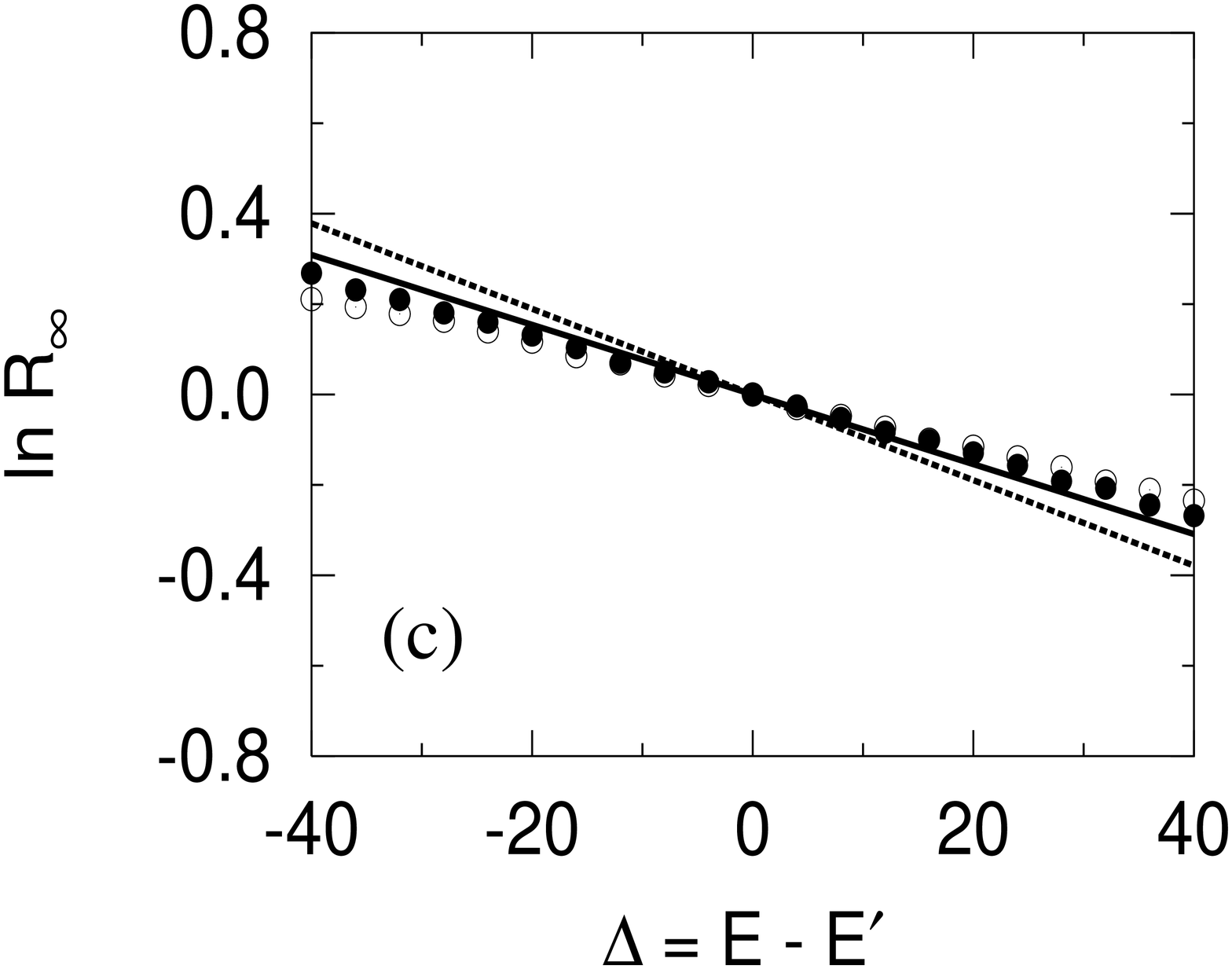, bbllx=20,bburx=630, bblly=100, bbury=750, angle=-90}}
	  }
    }
\end{center}
\par
\vspace{-6mm}
\caption{Normalized histograms for the driven system using 
heat bath (solid line) and Metropolis (dotted line) rates, at 
$\beta_{1}=1/3.550$ (a) and $\beta_{2}=1/3.650$ (b).
In (c), we show $\ln R_{\mathcal{\infty }}$ vs $E - E^{\prime}$.
Metropolis data are shown as open circles and are taken at
$\beta_{1}=1/3.200$ and $\beta_{2}=1/3.300$; the heat bath
data (filled circles) are taken at
$\beta_{1}^{\prime}=1/3.550$ and $\beta_{2}^{\prime}=1/3.650$. 
Two theoretical lines are shown: 
$-(\beta_{1}-\beta _{2})(E-E^{\prime })$ (dotted) and 
$-(\beta_{1}^{\prime} -\beta _{2}^{\prime})(E-E^{\prime })$
(solid). } 
\label{fig:histocomp2}
\end{figure}

\subsection{Histogram Ratio Analysis}

In the final section, we turn to a brief investigation of energy histograms.
Since Glauber and heat bath rates produce essentially identical data, we
restrict ourselves in the following to just heat bath and Metropolis rates.
To set the scene, we first show two histograms for the equilibrium system,
generated at, and slightly above, criticality: $T_{1}=2.269$ 
and $T_{2}=2.369$ (Figs.~7a and b, respectively). As
expected, the data for the different dynamics collapse very well, within
statistical errors. Not surprisingly, the peak position shifts to higher
energies with increasing temperature, while the width is largest at
criticality. In Fig.~7c, we plot the corresponding histogram ratio, 
Eq.~(\ref{eq:historatio}), and compare it to the predicted exponential 
form. The agreement is of course very good.

With Fig.~8, we enter novel territory. In analogy to the equilibrium plots,
Figs.~8a and b display the energy histograms of the driven system, at two
temperatures, $T_{1}=3.550$ and $T_{2}=3.650$,
for heat bath and Metropolis dynamics. The chosen temperatures correspond to
criticality and slightly above for heat bath rates; for Metropolis rates,
both are well inside the disordered phase. In contrast to the equilibrium
case, the histograms clearly depend on the choice of rates: the peak
positions are considerably higher for Metropolis than for heat bath rates.
At the same time, the width is largest for the system closest to
criticality, i.e., $T_{1}=3.550$ with heat bath rates.

A comment is in order, concerning the judicious choice of the two
temperatures which enter the histogram ratio. It applies to both the
equilibrium and the driven case. As we can see from Figs.~7 and 8, each
histogram displays a well-developed peak. Energies far away from the peak
position occur rarely, so that histograms are plagued by large statistical
errors in those regions. In order to yield a reliable ratio, the
corresponding histograms should overlap in their statistically meaningful
domains. Hence, the two chosen temperatures must not lie too far apart. 

In Fig.~8c, we present the histogram ratio for the driven system. For each 
dynamics, two temperatures close to their respective critical 
temperatures were chosen: $3.200$ and $3.300$ for Metropolis rates, 
and $3.550$ and $3.650$ for heat bath rates. Remarkably, we observe that 
the histogram ratio for both is again a
simple exponential, i.e., $\ln R_{\mathcal{\infty }}(E,E^{\prime })\propto
(E-E^{\prime })$, at least over the range shown. In stark contrast to the
equilibrium case, there is no a priori reason here to expect such behavior.
Instead, it indicates that $F(E,\beta )$ in Eq.~(\ref{dds-historatio})
depends sufficiently smoothly on $E$ as to allow an expansion in $\Delta
\equiv E-E^{\prime }$:

\begin{eqnarray}
\ln R_{\mathcal{\infty }}(E,E^{\prime }) &=&-\Delta \left[ \frac{\partial
F(E,\beta _{1})}{\partial E}-\frac{\partial F(E,\beta _{2})}{\partial E}%
\right] +O(\Delta ^{2})  \nonumber \\
&\equiv &-\alpha \Delta +O(\Delta ^{2})  \label{lin}
\end{eqnarray}

Hence, the slope of the data in Fig.~8c allows us to probe $\alpha $, 
as a function of temperature and dynamics. It manifestly differs 
from the equilibrium form $(\beta _{1}-\beta _{2})$. A more
systematic study is required to extract, and interpret, its
properties.


\section{\label{sec:level4}Conclusions}

We have simulated the equilibrium Ising lattice gas and its driven
non-equilibrium counterpart, using three different dynamics: Metropolis,
Glauber and heat bath. In the equilibrium case, all three rate functions
satisfy detailed balance with respect to the Ising Hamiltonian; as a
consequence, all stationary (time-independent) equilibrium quantities are
expected to be independent of the choice of the dynamics. Apart from
unavoidable statistical errors, our equilibrium data are of course perfectly
consistent with this expectation. For the driven system, this is no longer
the case: due to the drive, all three rate functions violate detailed
balance, and the `decoupling' of stationary properties from the chosen
dynamics no longer holds. Measuring two-point correlations and structure
factors in the disordered phase, we observe distinct
differences between the three dynamics. On the one hand, Metropolis rates lead
to a lower critical temperature, and hence generally weaker correlations in
the disordered phase, than either Glauber or heat bath rates. 
On the other hand, the
latter two generate practically indistinguishable data. These features can
be understood in terms of a few basic properties of the three rates: At a
given temperature, Metropolis rates tend to accept all 
moves with a somewhat higher probability than the other two rate functions. In
other words, a system which evolves under the Metropolis algorithm `sees' an
effectively higher temperature than if it were running under heat bath or
Glauber. A similar observation was made recently for field-driven Ising or
solid-on-solid interfaces, subject to Glauber and Metropolis dynamics:
there, Metropolis rates appear to lead to rougher interfaces and higher 
propagation velocities than Glauber rates \cite{PAR}. 
In contrast, heat bath and Glauber rates partition the unit
interval into the same subsections and accept/reject moves according to this
partition. As a result, they generate statistically indistinguishable
trajectories in configuration space, leading to essentially identical data.

It is essential to note, however, that the \emph{broad characteristics}, associated
with the breaking of detailed balance, are clearly observed in all three
dynamics: all structure factors show the typical discontinuity singularity
at the origin which, in turn, translates into power law decays of the
two-point correlation functions. To summarize, \emph{universal} features,
associated with \emph{global} symmetries, remain independent of \emph{local}
changes of dynamic rules, both near and far from equilibrium.

In a second part of this paper, we discuss the energy histograms $H(E,\beta
) $ associated with our two models, generated by heat bath and Metropolis
dynamics. For the equilibrium system, the independence of the choice of
dynamics is borne out again, while differences emerge in the driven
case. A simple ratio, $R(E,E^{\prime })$, constructed from two histograms
measured at different temperatures $\beta _{1}$ and $\beta _{2}$, allows us
to probe their functional form for a specified dynamics. In equilibrium,
the canonical distribution prescribes a simple exponential dependence, $\ln
R_{0}=-(\beta _{1}-\beta _{2})(E-E^{\prime })$. Remarkably, its
non-equilibrium counterpart $\ln R_{\infty }$ is \emph{also} exponential in 
$(E-E^{\prime })$. This behavior indicates a smooth dependence of $F(E,\beta
)\equiv -\ln H(E,\beta )$ on $E$, allowing us to linearize 
$\ln R_{\mathcal{\infty }}$ in $(E-E^{\prime })$. 
The slope of the resulting straight line
depends on the dynamics and probes the derivative $(\partial F/\partial E)$.
Further, and more detailed, studies of this type may reveal some of the
hidden ``thermodynamics'' of this remarkably complex non-equilibrium steady
state.

\section{\label{sec:level5}Acknowledgment}

\vspace{-1mm} 
We thank Royce K.P. Zia and Per A. Rikvold for fruitful discussions. This research
was supported in part by National Science Foundation Grants No. DMF-0094422
and DMR-0088451.

\bigskip \bigskip \newpage

\newpage 

\end{document}